\documentclass[a4paper,11pt]{article}
\pdfoutput=1
\usepackage{jheppub}
\usepackage[T1]{fontenc}

\usepackage{amsmath,amssymb,amsthm,amscd}
\usepackage{tikz} 
\usetikzlibrary{decorations.pathmorphing,calc}
\usetikzlibrary{calc}
\tikzset{node distance=2cm, auto}
\usepackage{epsfig}
\usepackage{psfrag,comment}
\usepackage{graphicx}

\title{Generalized phase transitions in Lovelock gravity}

\author[a]{Xi\'an~O.~Camanho}
\affiliation[a]{Max-Planck-Institut f\"ur Gravitationsphysik, Albert-Einstein-Institut, 14476 Golm, Germany}
\emailAdd{xian.camanho@aei.mpg.de}

\author[b,c]{Jos\'e~D.~Edelstein,}
\affiliation[b]{Department of Particle Physics and IGFAE, University of Santiago de Compostela, E-15782 Santiago de Compostela, Spain}
\affiliation[c]{Centro de Estudios Cient\'\i ficos, CECS, Casilla 1469, Valdivia, Chile}
\emailAdd{jose.edelstein@usc.es}

\author[d,e]{Gast\'on~Giribet}
\affiliation[d]{University of Buenos Aires FCEN-UBA and IFIBA-CONICET, Ciudad Universitaria, Pabell\'on I, 1428, Buenos Aires, Argentina}
\affiliation[e]{Abdus Salam ICTP, Strada Costiera 11, 34100, Trieste, Italy}
\emailAdd{gaston@df.uba.ar}

\author[f]{Andr\'es~Gomberoff}
\affiliation[f]{Universidad Andres Bello, Departamento de Ciencias F\'\i sicas, Av. Rep\'ublica 252, Santiago, Chile}
\emailAdd{agomberoff@unab.cl}

\abstract{
We investigate a novel mechanism for phase transitions that is a distinctive feature of higher-curvature gravity theories. For definiteness, we bound ourselves to the case of Lovelock gravities. These theories are known to have several branches of asymptotically AdS solutions. Here, extending our previous work, we show that phase transitions among some of these branches are driven by a thermalon configuration: a bubble separating two regions of different effective cosmological constants, generically hosting a black hole in the interior. Above some critical temperature, this thermalon configuration is preferred with respect to the finite-temperature AdS space, triggering a sophisticated version of the Hawking-Page transition. After being created, the unstable bubble configuration can in general dynamically change the asymptotic cosmological constant. While this phenomenon already occurs in the case of a gravity action with square curvature terms, we point out that in the case of Lovelock theory with cubic (and higher) terms new effects appear. For instance, the theory may admit more than one type of bubble and branches that are in principle free of pathologies may also decay through the thermalon mechanism. We also find ranges of the gravitational couplings for which the theory becomes sick. These add up to previously found restrictions to impose tighter constraints on higher-curvature gravities. The results of this paper point to an intricate phase diagram which might accommodate similarly rich behavior in the dual CFT side.
}


\subheader{~\hfill ${\rm AEI}-2013-267$}

\begin{document}

\maketitle

\allowdisplaybreaks

\section{Introduction}

Phase transitions between two competing vacua of a given theory are a quite common phenomenon in physics. They occur when some parameter of the system is varied so that the (free) energy of the actual vacuum becomes greater than the other. If the energy barrier between the two is big enough, the system may stay in the false vacuum for some time (metastability), and then decay via quantum tunneling or, at finite temperature, jumping over the wall due to a thermal kick. The decay of metastable systems usually proceed by nucleation of bubbles of true vacuum inside the false vacuum. In field theories at zero temperature, this was first studied by Coleman in his classic paper \cite{Coleman1977}. There, he introduced Euclidean methods for computing the probability of the quantum nucleation of a bubble, whose dynamics, after nucleation, may be followed classically. In the first (tree-level) semiclassical approximation, the probability of bubble nucleation is given by
\begin{equation}
P \propto e^{-\mathcal{I}_{\rm E}} ~,
\label{prob}
\end{equation}
where $\mathcal{I}_{\rm E}$ is the Euclidean action of the system evaluated at the appropriate solution; in this case, the instanton. It is a time-dependent solution which, in the simplest case of a particle in a potential, starts and ends its trajectory at the bottom of the false vacuum (given that the potential in the Euclidean section is the negative of its Lorentzian counterpart, it is a local maximum).
\begin{figure}[ht]
\centering
\includegraphics[width=12cm]{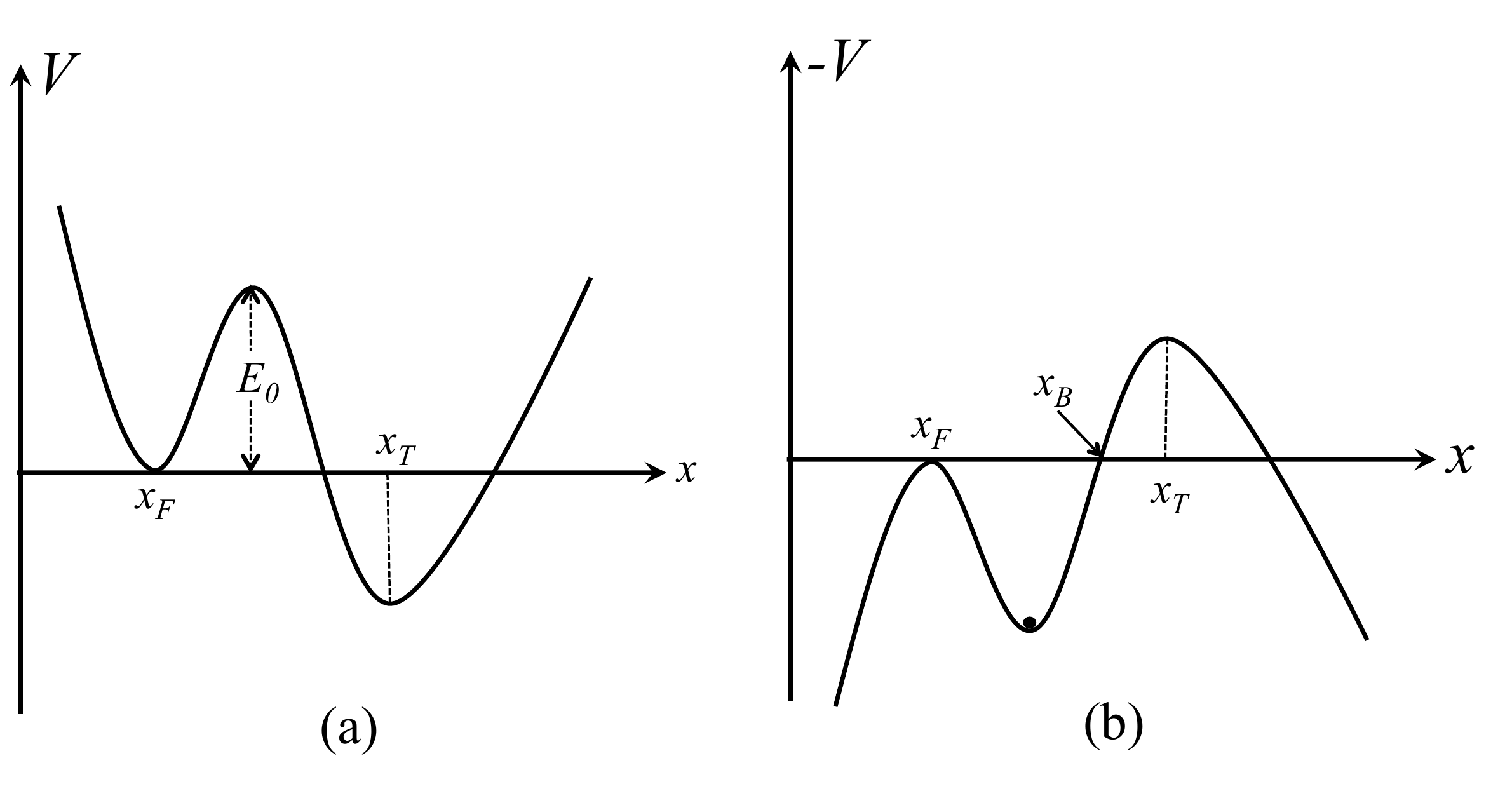}
\caption{In (a) we depict a particle in a potential with two local minima. At $x_F$, the {\it false vacuum}, while the true vacuum is at $x=x_T$. There is a potential barrier of height $E_0$ between the two vacua. In (b) the Euclidean counterpart of the same system. The potential is now $-V$.}
\label{pot}
\end{figure}
This is the point $x=x_F$ in Figure \ref{pot}(b), where the (system) particle starts its trajectory, then bounces at $x=x_B$, and finally gets back to $x_F$ in infinite time. The work of Coleman generalizes this mechanism to a scalar field theory. The instanton corresponds to  a scalar field configuration with  SO(4) symmetry.

The technique was then generalized by Linde, who considered a scalar field at finite temperature \cite{Linde1981}. In this case, the probability of nucleation is still given by (\ref{prob}), but the Euclidean action has to be evaluated using a different classical configuration, which has  SO(3) symmetry. In the mechanical example of Figure \ref{pot}, if the temperature is not high enough, this is the solution oscillating inside the Euclidean well, between two points in the interval $(x_F,x_B)$. The period  $\beta$ of that solution is identified with the inverse temperature, $\beta=1/T$. This period has a minimum for small oscillations at the bottom of the well. For temperatures higher than that one may use the static solution with the particle at the bottom, which of course has any periodicity. It is easy to see that for this case, (\ref{prob}) gives precisely the Boltzmann factor $e^{-\beta E_0}$ that one expects for the probability of the particle to jump over the barrier. This is the {\it sphaleron} or {\it thermalon} \cite{Klinkhamer1984,Gomberoff2004}. 

Gravitational instantons where first discussed by Coleman and de Luccia in \cite{Coleman1980}, where a scalar field with a potential interacts with a dynamical metric. Now, the different vacua correspond to solutions with different cosmological constants. The false vacuum decays by nucleating an expanding bubble of true vacuum. Later, Brown and Teitelboim \cite{Brown1987,Brown1988}  found an analog instanton when gravity was coupled to an electromagnetic $3$-form potential and its sources, electrically charged membranes. In that case, there are infinitely many false vacua, and the decay may proceed many times, changing the (positive) cosmological constant at each step. The authors showed that this mechanism could relax the cosmological term, providing a possible mechanism for understanding the {\it cosmological constant problem} \cite{Weinberg1989}. For finite temperatures, the physical system may also decay. Now a thermalon solution controls the decay rate, and, interestingly enough, the decay of a pure de Sitter geometry ends up leaving a black hole behind \cite{Gomberoff2004}.

In this paper we show an analog process that occurs in higher-curvature theories of gravity. In general, these theories contain degenerate vacua even in the absence of matter. Furthermore, one vacuum may decay into the other by nucleating bubbles made of nothing but gravity itself. This phenomenon has been studied in \cite{Camanho2012,Camanho2013a} for the special case of Lanczos-Gauss-Bonnet (LGB) quadratic deformation of Einstein's gravity \cite{Lanczos}, where it was shown that a sophisticated version of the Hawking-Page phase transition occurs for any number of dimensions and for solutions with different horizon topology. This transition connects in a dynamical fashion asymptotically AdS geometries with different effective cosmological constants. The quadratic case, though, is {\it a priori} a sick example, given that one of the vacua necessarily suffers from the so-called Boulware-Deser (BD) instability \cite{Boulware1985a}. Here, we would like to remove the splinter and show that the phase transition is not at all due to (or caused by) the unhealthy nature of the BD vacuum, but rather a distinctive feature of higher-curvature models. Furthermore, we will see that in the case of higher-order corrections to the gravity action ({\it e.g.} cubic Lovelock theory) novel features may happen, such as the existence of different species of bubbles that can nucleate in an unstable AdS vacuum, with different probabilities.

The content of the paper is organized as follows. In the next Section we will review some basics of Lovelock gravity \cite{Lovelock1971} that are necessary for the computations of the paper, making particular emphasis on those aspects that are less abundant in the literature. We focus on Lovelock theory due to its distinctive healthy feature of having equations of motion that do not involve terms with more than two derivatives of the metric in any space-time dimensions. In Section 3 we provide a one-dimensional example, the higher order free particle, with the aim of introducing some peculiar aspects of Lovelock theory in a much simpler setup. We devote Section 4 to the derivation of the junction conditions necessary to construct the thermalon. We then switch to Lorentzian signature to explore the bubble dynamics. This leads to a number of issues such as the thermalon existence and stability, the fate of the bubbles in relation to the advocated phase transitions, and the status of the cosmic censorship hypothesis under this process.

In Section 5 we elaborate on the thermodynamics of the process, performing some consistency checks. The general analysis is difficult to pursue analytically. We go as far as we can and specialize in the LGB case, when necessary, for definiteness. These results allow us to finally address the issue of the generalized Hawking-Page transitions in Lovelock gravity in Section 6. For planar topology, formulas simplify drastically and we can explore the general case. We focus on the cubic theory to illustrate some of the new features that arise in comparison to the LGB gravity. For non-planar topologies we are forced to stick ourselves to the LGB theory to obtain concrete analytic results. We conclude this article in Section 7, where we summarize our results, comment on their interpretation in the context of the gauge/gravity duality, and speculate on other prospective problems that higher-curvature gravities might entail. Finally, we include a short Appendix making an important remark on causality violation in AdS vacua (or black hole solutions, whenever they exist) corresponding to higher-curvature branches.

\section{Lovelock basics \label{Lovelock}}

Let us first review some basics facts of Lovelock theory that are needed in this article. For a more complete and detailed account, see \cite{GarraffoG,CamanhoE3,Edelstein,Josinho}. In $d$ dimensions, the action can be written in terms of the vierbein one-form, $e^{a}=e_{\mu}^{a}\,dx^{\mu}$, and the Levi--Civita spin connection, $\omega^{ab} = \omega^{ab}_\mu\,dx^{\mu}$, as a sum of $K + 1$ terms, $K\leq \left[ (d-1)/2 \right]$, consisting of bulk and boundary contributions,
\begin{equation}
\mathcal{I} = \sum_{k=0}^{K}{\frac{c_{k}}{d-2k}}\,\left( \int_{\mathcal{M}}\!\!\! \mathcal{L}_k - \int_{\partial\mathcal{M}}\!\!\! \mathcal{Q}_k \right) ~.
\label{LLaction}
\end{equation}
where $\mathcal{M}$ is the space-time manifold and $\partial\mathcal{M}$ its boundary. Both fields $e^{a}$ and $\omega^{ab}$ should be {\it a priori} considered equally fundamental. We will however consider a sector of this theory characterized by the vanishing torsion condition, $T^a := de^{a} + \omega_{~b}^{a} \wedge e^{b} = 0$, which automatically fulfills the equations of motion resulting from the variation of (\ref{LLaction}) with respect to the spin connection.

The coefficients $c_{k}$ are coupling constants with length dimensions $L^{2(k-1)}$. The $k^{\text{th}}$-bulk contribution can be written as \cite{Lovelock1971}
\begin{equation}
\mathcal{L}_k = R^{a_{1}a_{2}} \wedge \cdots \wedge R^{a_{2k-1}a_{2k}} \wedge \mathfrak{E}^{(d)}_{a_{1} \cdots a_{2k}} ~,
\label{LLbulk}
\end{equation}
in terms of the curvature 2-form, $R^{ab} = d \omega^{ab} + \omega_{~c}^{a} \wedge \omega^{cb}$, and a $(d - 2k)$-form constructed from the vierbein and the antisymmetric symbol,
\begin{equation}
\mathfrak{E}^{(d)}_{a_{1} \cdots a_{2k}} = \epsilon_{a_{1} \cdots a_{d}}\, e^{a_{2k+1}} \wedge \cdots \wedge e^{a_{d}} ~;
\label{gothicEform}
\end{equation}
the subindices and superindices refer to components in the tangent bundle. Each term in (\ref{LLaction}) corresponds to the expression of the Euler characteristic in $2k$ dimensions extended to $d$ dimensions; the $k^{\text{th}}$-boundary piece, in turn, reads \cite{Myers1987}
\begin{equation}
\mathcal{Q}_k = k \int_0^1\!\! d\xi \ \theta^{a_1 a_2} \wedge \mathfrak{F}^{a_3 a_4}(\xi) \wedge \cdots \wedge \mathfrak{F}^{a_{2k-1} a_{2k}}(\xi) \wedge \mathfrak{E}^{(d)}_{a_1 \cdots a_{2k}} ~,
\label{LLboundary}
\end{equation}
where $\theta^{ab}$ is the second fundamental form associated to the extrinsic curvature, and
\begin{equation}
\mathfrak{F}^{ab}(\xi) \equiv R^{ab} + (\xi^2 - 1)\, \theta_{~e}^{a} \wedge \theta^{eb} ~.
\label{curlyF}
\end{equation}
These boundary terms naturally appear in the Chern-Weil generalization of the Gauss-Bonnet theorem to manifolds with boundaries.

Notice that $\theta^{ab} = \omega^{ab} - \omega_0^{ab}$ is covariant under changes of frame and has only normal components on $\partial\mathcal{M}$, since $\omega_0^{ab}$ is the spin connection adapted to the product metric on $\mathcal{M}$ (see \cite{Eguchi1980} for details). We can make use of this splitting to write 
\begin{equation}
R^{ab} = R_0^{ab} + D_0\theta^{ab} + \theta_{~c}^{a} \wedge \theta^{cb} ~,
\label{Rdecompose}
\end{equation}
where $D_0\theta^{ab} = d\theta^{ab} + {\omega_0}_{~c}^{a} \wedge \theta^{cb} + \theta_{~c}^{a} \wedge {\omega_0}^{cb}$, and $R_0^{ab}$ is the intrinsic curvature at the boundary. Both $\theta^{ab}$ and $D_0\theta^{ab}$ are zero unless one of the indices is along the normal direction. Therefore, taking into account that there is a $\theta^{ab}$ factor in eq.(\ref{LLboundary}),
\begin{equation}
\mathcal{Q}_k = k \int_0^1\!\! d\xi \ \theta^{a_1 a_2} \wedge \mathfrak{F}_0^{a_3 a_4}(\xi) \wedge \cdots \wedge \mathfrak{F}_0^{a_{2k-1} a_{2k}}(\xi) \wedge \mathfrak{E}^{(d)}_{a_1 \cdots a_{2k}} ~,
\label{LLboundary-bis}
\end{equation}
where $\mathfrak{F}_0^{ab}(\xi) \equiv R_0^{ab} + \xi^2\, \theta_{~e}^{a} \wedge \theta^{eb}$, which is nothing but the value of $\mathfrak{F}^{ab}(\xi)$ when both indices are tangential. Let us be more explicit in relating $\theta^{ab}$ to the extrinsic curvature. They are related by the following expression
\begin{equation}
\theta^{ab} = n^a K^b - n^b K^a ~,
\label{extrinsiK}
\end{equation}
where $n^a$ is the unit normal vector at $\partial\mathcal{M}$, and $K^a = K^a_{~b}\, e^b$ is the extrinsic curvature 1-form. Given that $\theta_{~c}^{a} \wedge \theta^{cb} = - K^a \wedge K^b$, the boundary terms (\ref{LLboundary}) can be written in terms of the intrinsic and extrinsic curvatures of the boundary,
\begin{equation}
\mathcal{Q}_k = -2k \int_0^1\!\! d\xi \ K^{a_1} \wedge \mathfrak{F}_0^{a_2 a_3}(\xi) \wedge \cdots \wedge \mathfrak{F}_0^{a_{2k-2} a_{2k-1}}(\xi) \wedge \mathfrak{E}^{(d)}_{a_1 \cdots a_{2k-1}} ~,
\label{LLboundaryK}
\end{equation}
where, now, $\mathfrak{F}_0^{ab}(\xi) \equiv R_0^{ab} - \xi^2\, K^{a} \wedge K^{b}$.

Despite the unconventional appearance of the precedent formulas, notice that the zero$^{\text{th}}$ contribution in (\ref{LLaction}) is nothing but the cosmological constant term (we set $2\Lambda = -(d-1)(d-2)/L^{2}$, {\it i.e.}, $c_0 = 1/L^2$), with vanishing\footnote{In the context of holographic renormalization, it is sometimes necessary to introduce a constant $\mathcal{Q}_0$ --{\it i.e.}, a boundary cosmological term-- in order to regularize otherwise divergent quantities.} $\mathcal{Q}_0$. The first term, in turn, gives raise to the Einstein--Hilbert (EH) action in the bulk, and the Gibbons-Hawking term \cite{GibbonsHawking} at the boundary (we normalize the Newton constant to $16\pi (d-3)!\,G_{N}=1$, {\it i.e.}, $c_1 = 1$). As is well-known, the latter is necessary for the variational principle to be well defined. This is actually the case for all boundary contributions given by $\mathcal{Q}_k$. The second term, quadratic in the Riemann curvature, is the LGB action (we take $c_2 = \lambda L^{2}$, and call $\lambda$ the LGB {coupling}).

Due to the non-linearity of the equations of motion, these theories generally admit more than one maximally symmetric solution, $R_{\mu \nu \alpha\beta} = \Lambda_i (g_{\mu\alpha}\,g_{\nu\beta} - g_{\mu\beta}\,g_{\nu\alpha})$; (A)dS vacua with effective cosmological constants $\Lambda_{i}$, whose values are determined by (the real roots of) a polynomial equation \cite{Boulware1985a}, 
\begin{equation}
\Upsilon[\Lambda] \equiv \sum_{k=0}^{K}c_{k}\,\Lambda^{k} = c_{K}\prod_{i=1}^{K}\left( \Lambda -\Lambda _{i}\right) = 0 ~.
\end{equation}
$K$ being the highest power of the curvature (without derivatives) in the field equations. The dynamics of perturbative excitations about a given vacuum, say $\Lambda_\star$, is that of Einstein's theory with an effective Newton constant whose sign is that of $\Upsilon'[\Lambda_\star]$. This is the source of BD instabilities and, interestingly enough, is holographically related to a unitarity condition in the dual CFT \cite{Camanho2010d}.

Any vacua is {\it a priori} suitable in order to define boundary conditions for the gravity theory we are interested in; {\it i.e.}, we can define sectors of the theory as classes of solutions that asymptote to a given vacuum \cite{CamanhoE3}. In that way, each branch has associated static solutions, representing either black holes or naked singularities with the same asymptotics,
\begin{equation}
ds^{2}=-f(r)\,dt^{2}+\frac{dr^{2}}{h(r)}+r^{2}\ d\Sigma_{d-2,\sigma}^{2} ~, \qquad \qquad f,h \xrightarrow{r\rightarrow \infty} -\Lambda_i r^2 ~,
\label{bh2}
\end{equation}
where
$$
d\Sigma_{d-2,\sigma}^{2} = \frac{d\rho^2}{1 - \sigma \rho^2/L^2}+\rho^2 d\Omega^2_{d-3} ~,
$$
is the metric of a $(d-2)$-dimensional manifold of negative, zero or positive constant curvature ($\sigma =-1, 0, 1$ parameterizing the different horizon topologies), and $d\Omega^2_{d-3}$ is the round metric of the unit $(d-3)$-sphere.

The equations of motion reduce to a single first order differential equation for $f = h$, that can be easily solved in terms of $g=(\sigma - f)/r^{2}$ as \cite{Wheeler1986,Wheeler1986a}
\begin{equation}
\Upsilon \lbrack g] = \sum_{k=0}^{K}c_{k}\,g^{k} = \frac{M}{r^{d-1}} ~,
\label{eqg}
\end{equation}
with a single integration constant $M$ that is interpreted as the mass.\footnote{For convenience, we omit a normalization factor $((d-2)!\,V_{d-2})^{-1}$, where $V_{d-2}$ is the volume of the unit radius constant curvature manifold whose line element is $d\Sigma_{d-2,\sigma}^{2}$ (see \cite{CamanhoE3} for details).} There are in general $K$ branches (monotonic sections) of this implicit polynomial solution, and they are in one-to-one correspondence with the classes discussed above \cite{CamanhoE3}.

The main motivation of the present work is that of studying transitions between different branches of solutions, like the one presented in \cite{Camanho2012}. This is important in order to investigate whether a new type of instability involving non-perturbative solutions occurs in the theory.

\section{Higher order free particle}

The existence of branch transitions in higher-curvature gravity theories is a concrete expression of the multivaluedness features that appear in different dynamical problems, and in particular in these theories. In general, in Lovelock theory, the canonical momenta, $\pi_{ij}$, are not invertible functions of the velocities, $\dot{g}^{ij}$ \cite{Teitelboim1987}. An analogous situation may be illustrated by means of a much simpler one-dimensional example \cite{Henneaux1987b}. Consider a free particle Lagrangian containing higher powers of velocities,
\begin{equation}
L(\dot{x}) = \frac{1}{2} \dot{x}^2 - \frac13 \dot{x}^3 + \frac1{17} \dot{x}^4 ~.
\label{paction}
\end{equation}
In the Hamiltonian formulation, the equation of motion just implies the constancy of the conjugate momentum, $\dot{p}=0$. However, this being multivalued (also the Hamiltonian), the solution is not unique. Fixing boundary conditions $x(t_{1,2})=x_{1,2}$, an obvious solution would be constant speed $\dot{x}=(x_2-x_1)/(t_2-t_1)\equiv v$, but we also have jumping solutions with constant momentum and the same mean velocity. Obviously, for that to happen at least one of the degenerate velocities has to be bigger than $v$ and one smaller.

\begin{figure}[ht]
\centering
\includegraphics[width=10.3cm]{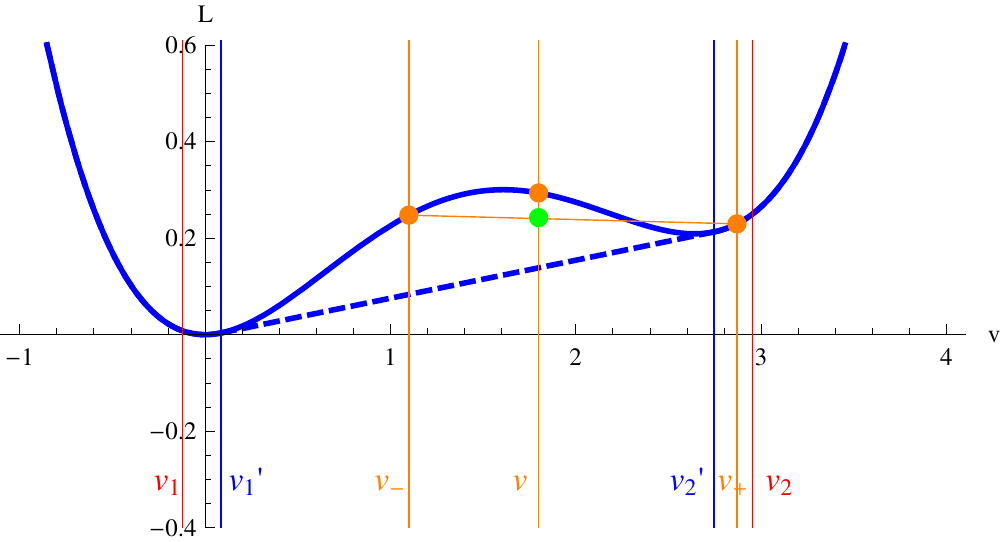}  \\
\includegraphics[width=10.3cm]{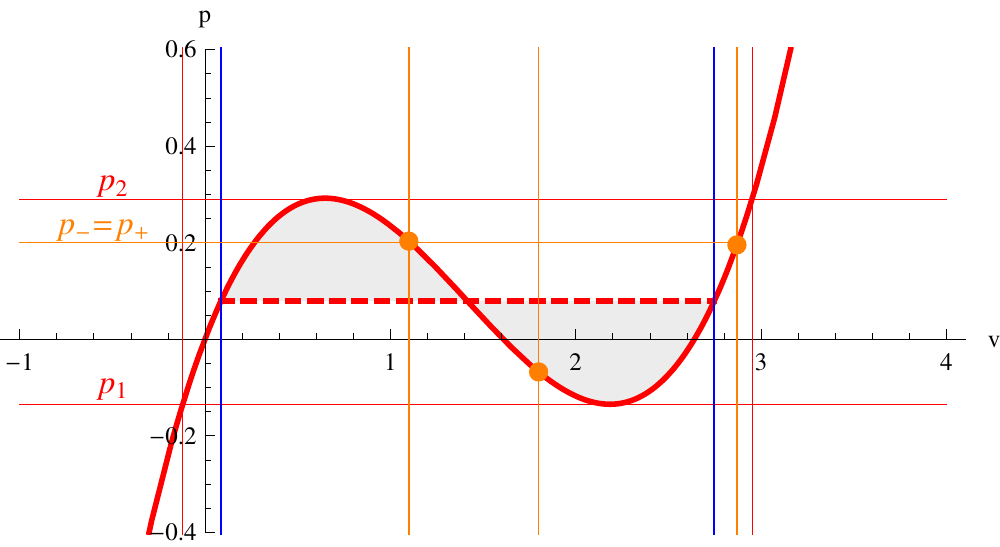} 
\caption{Lagrangian and momentum for the action (\ref{paction}). For the same mean velocity $v$, the action is lower for jumps between $v_\pm$ (green dot, light gray in b$\&$w) than for constant speed, the minimum action corresponding to the value on the dashed line ({\it effective} Lagrangian).}
\label{L-v}
\end{figure}

\begin{figure}[ht]
\centering
\includegraphics[width=10.3cm]{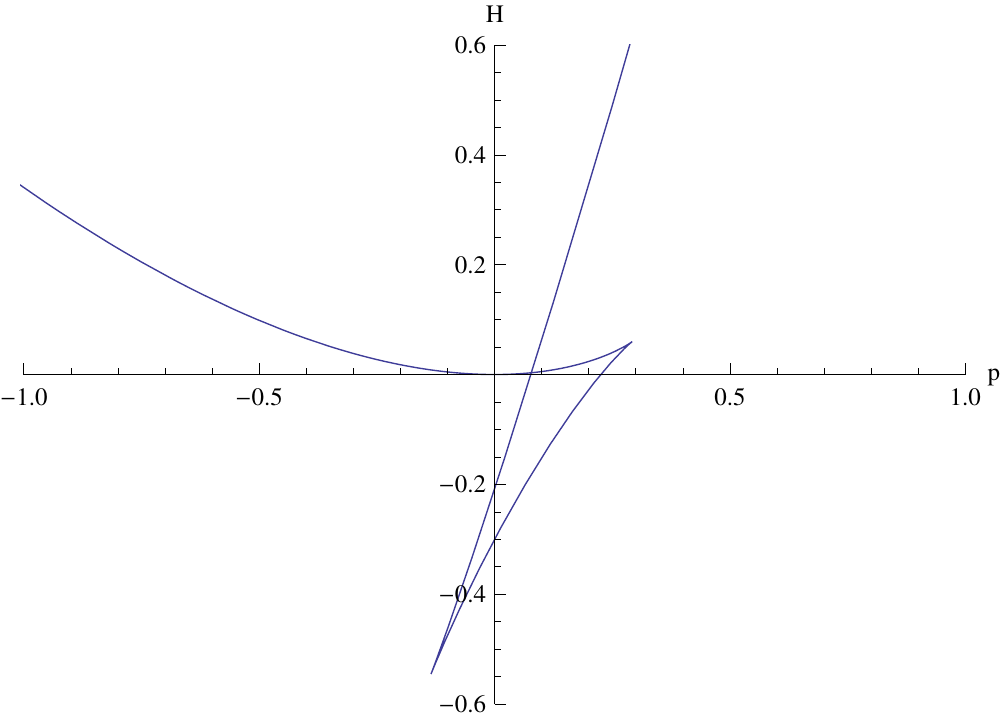} 
\caption{Hamiltonian as a function of the momentum for the action (\ref{paction}). Multivalued momenta correspond to the swallowtail part of the curve whereas those corresponding to the {\it effective} Lagrangian are the ones without that part of the curve, just the two upper branches. The crossing correspond to the {jump} depicted in Figure \ref{L-v} as a dashed line}
\label{H-p}
\end{figure}

In the example given in (\ref{paction}), for mean velocities in the range $(v_1,v_2)$ --that corresponds to multivalued momentum-- (see Figure \ref{L-v}), the solutions are infinitely degenerate as the jumps may occur at any time and unboundedly in number, as long as the mean velocity remains the same.
This degeneracy is lifted however when the value of the action is taken into account \cite{ChiHe}. The minimal action path is the na\"ive one for mean velocities outside the range $(v_1',v_2')$, whereas in that range it corresponds to arbitrary jumps between the two extremal velocities. One can actually compare the action for both kinds of trajectories directly in the figure, where we have depicted one example. In the top figure, we have plotted the Lagrangian as a function of the velocity. When the latter is constant, $\dot{x}=v$, the action will just be the Lagrangian $L(v)$ multiplied by the time span of the trajectory. On the other hand, for a jumping trajectory, the mean Lagrangian can be found as the intersect of the straight line joining the values of the Lagrangian for both values of the velocity, $\dot{x}=v_\pm$, and that of $\dot{x}=v$,
\begin{equation}
\bar{L}(v;v_\pm) = L(v_-) + \frac{v-v_-}{v_+-v_-} L(v_+) ~.
\end{equation}
Notice that the action corresponding to $v$ is bigger than the one corresponding to jumps between $v_{\pm}$ (green dot). This means that the Lagrangian is minimal for the lowest lying such straight line and it is easy to see that it corresponds to the dashed line for mean velocities, $v_1'<v<v_2'$. Outside that range, the momentum is a convex function of the velocity and any line joining two points on opposite sides of a given velocity will necessarily yield a higher action. The effective Lagrangian (dashed line) is actually also a convex function of the velocities, and the effective momentum dependence corresponds to the Maxwell construction. In that case, the dashed line corresponds to jumps conserving both the energy and the momentum, {\em i.e.}, the crossing point on Figure \ref{H-p}.
Furthermore, the lowest energy state does not correspond to $\dot{x}=0$, even classically. This phenomenon has been referred to, quite recently, as a {\it time crystal} \cite{Shapere2012,Wilczek2012}. The quantum mechanical version of the model is well defined \cite{Henneaux1987b}.

In the presence of a potential, the na\"ive choice of continuous velocities runs into problems as we would hit the degenerate points $\frac{dp}{d\dot{x}}=0$, where the acceleration is not well defined. In the above effective approach, however, the momentum is a monotonous function of the velocity except at the {\it transition} which corresponds to conservation not only of the momentum, but of the energy.

\section{Generalized Hawking-Page transitions}

In the context of General Relativity in asymptotically AdS space-times, the Hawking-Page phase transition \cite{HP} is the realization that above certain temperature the dominant saddle in the gravitational partition function comes from a black hole, whereas for lower temperatures it corresponds to the thermal vacuum. The {\it classical} solution is the one with least Euclidean action among those with a smooth Euclidean section. In the case of Lovelock gravities the same phenomenon can be observed to occur \cite{CamanhoE3}.

When one deals with higher-curvature gravity, however, there is a crucial difference that has been overlooked in the literature. In addition to the usual continuous and differentiable metrics (\ref{bh2}), one may construct distributional metrics by gluing two solutions corresponding to different branches across a spherical shell or {\it bubble} \cite{wormholes,Garraffo2008b}. The resulting configuration will be continuous at the bubble --with discontinuous derivatives, even in absence of matter: the higher-curvature terms can be thought of as a sort of matter source for the Einstein tensor--. The existence of such {\it jump} metrics, as for the one-dimensional example of the previous Section, is also due to the multivaluedness of momenta in the theory. In the context of gravitational theories, continuity of momenta is equivalent to the junction conditions that need to be imposed at the location of the bubble. In the EH case, these are nothing but Israel's junction conditions \cite{Israel1967}. Being linear in the velocities, they imply the continuity of the derivatives of the metric. The generalization of these conditions for higher-curvature gravity contain higher powers of velocities, thus allowing for more general situations.

Static bubble configurations, when they exist, have a smooth Euclidean continuation. It is then possible to calculate the value of the action and compare it to all other solutions with the same asymptotics and temperature. This analysis has been performed for the LGB action \cite{Camanho2012} for unstable boundary conditions \cite{Boulware1985a}. The result suggests a possible resolution of the instability through bubble nucleation. The phenomenon is quite general, as we will see in the present work. It presumably holds for more general classes of theories. One may even think of the possibility of having different gravity theories on different sides of the bubble. This has a nice physical interpretation if we consider the higher order terms as sourced by other fields that vary across the bubble. For masses above $m^2>|\Lambda_{\pm}|$ a bubble made of these fields will be well approximated by a thin wall and we may integrate out the fields for the purpose of discussing the thermodynamics. If those fields have several possible vacuum expectation values leading to different theories we may construct interpolating solutions in essentially the same way discussed in this paper. In this case the energy carried by the bubble can likely be interpreted as the energy of the fields we have integrated out. 

\subsection{Junction conditions}

We are interested in configurations involving a timelike (here $n^2=1$) junction surface dividing two regions, each of them given by a solution corresponding to a different branch of the Lovelock action. To this end, we will split the geometry as $\mathcal{M} = \mathcal{M}_-\, \cup\, ( \Sigma \times \xi )\, \cup\, \mathcal{M}_+$, where $\mathcal{M}_\mp$  denote the interior/exterior regions respectively, $\Sigma$ is the codimension one junction (hyper)surface, and $\xi \in [0,1]$ is a real segment/parameter used to interpolate both regions.\footnote{See \cite{Gravanis2003} for a related construction.} The matching condition can be obtained from an action principle constructed by means of two auxiliary quantities \cite{Gravanis2009}
\begin{equation}
E^a = \xi\, e^a_+ + (1-\xi)\, e^a_- ~, \qquad \Omega^{ab} = \xi\, \omega^{ab}_+ + (1-\xi)\, \omega^{ab}_- ~,
\end{equation}
and the associated generalized curvatures
\begin{eqnarray}
\mathcal{R}^{ab} & = & d \Omega^{ab} + \Omega_{~c}^{a} \wedge \Omega^{cb} \nonumber \\ [0.5em]
& = & \xi\, R^{ab}_+ + (1-\xi)\, R^{ab}_- - \xi (1-\xi)\, ({\omega_+} - {\omega_-})_{~c}^{a} \wedge ({\omega_+} - {\omega_-})^{cb} ~,
\label{genR}
\end{eqnarray}
and
\begin{equation}\mathfrak{F}^{ab} = \delta\Omega^{ab} + \mathcal{R}^{ab} = d\xi \wedge (\omega_+ - \omega_-)^{ab} + \mathcal{R}^{ab} ~,
\end{equation}
where $\delta$ is the exterior derivative on the convex simplex in the space of connections, $\delta = d\xi\,\frac{\partial}{\partial \xi}$. These curvatures are used to construct the secondary characteristic classes \cite{ChernSimons}. Their use in the context of Chern--Simons gravities has been considered in \cite{Mora2006}.

In the bulk regions, $\xi$ takes a fixed value ($\xi=0,1$ respectively in $\mathcal{M}_\mp$), whereas it runs from 0 to 1 in $\Sigma$ and has to be integrated over. The so-called {\it secondary} action is obtained by substituting $e^a$ by $E^a$ and $R^{ab}$ by $\mathfrak{F}^{ab}$ in (\ref{LLbulk}),
\begin{equation}
\widetilde{\mathcal{L}}_k = \mathfrak{F}^{a_{1}a_{2}} \wedge \cdots \wedge \mathfrak{F}^{a_{2k-1}a_{2k}} \wedge \mathfrak{E}^{(d)}_{a_{1} \cdots a_{2k}} ~,
\label{LLbulkjunc}
\end{equation}
where the replacement $e^a$ by $E^a$ is also understood in $\mathfrak{E}^{(d)}_{a_{1} \cdots a_{2k}}$.

We can readily expand the secondary action in powers of $\delta\Omega$, taking into account that $\delta\Omega \wedge \delta\Omega = 0$. The leading term corresponds to the bulk integrals on $\mathcal{M}_\mp$ and contributes to the standard Lovelock action (\ref{LLaction}), while the first order term captures the integral along $\xi$ on $\Sigma$, 
\begin{equation}
\widetilde{\mathcal{Q}}_k = - k \int_0^1\! d\xi\ (\omega_+ - \omega_-)^{a_{1}a_{2}} \wedge \mathcal{R}^{a_{3}a_{4}} \wedge \cdots \wedge \mathcal{R}^{a_{2k-1}a_{2k}} \wedge \mathfrak{E}^{(d)}_{a_{1} \cdots a_{2k}} ~.
\label{matchaction}
\end{equation}
If we further impose the continuity of the metric such that $E^a=e^a$ for some choice of vielbein, we may also define $\theta^{ab}_\pm = \omega^{ab}_\pm - \omega^{ab}_0$. Thereby, $\omega^{ab}_+ - \omega^{ab}_-$ can be replaced by $\theta^{ab}_+ - \theta^{ab}_-$, which is zero unless one of the indices lies along the normal direction. Thus, the only relevant contributions to $\mathcal{R}^{ab}$ in (\ref{matchaction}) are those whose two indices are tangent to $\Sigma$. Now,
\begin{equation}
\mathcal{R}^{ab} = R_0^{ab} + {\Theta}_{~c}^{a} \wedge {\Theta}^{cb} + \ldots ~,
\label{genRd}
\end{equation}
where $R_0^{ab}$ is the intrinsic curvature of $\Sigma$ computed from $\omega_0^{ab}$, ${\Theta}^{ab} \equiv \xi\, {\theta}_+^{ab} + (1-\xi)\, {\theta}_-^{ab}$, and the dots amount to terms having at least one index in the normal direction. Recalling the relation between the second fundamental form and the extrinsic curvature (\ref{extrinsiK}), we can finally write
\begin{equation}
\widetilde{\mathcal{Q}}_k =  2 k \int_0^1\! d\xi\ (K_+ - K_-)^{a_1} \wedge \mathcal{R}_{\rm j}^{a_{2}a_{3}} \wedge \cdots \wedge \mathcal{R}_{\rm j}^{a_{2k-2}a_{2k-1}} \wedge \mathfrak{E}^{(d)}_{a_{1} \cdots a_{2k-1}} ~,
\label{matchactionbis}
\end{equation}
where $\mathcal{R}_{\rm j}^{ab}(\xi) \equiv R_0^{ab} - (\xi\, K_+ + (1-\xi)\, K_-)^a \wedge (\xi\, K_+ + (1-\xi)\, K_-)^b$. Strikingly enough, we can split $\widetilde{\mathcal{Q}}_k = \mathcal{Q}^{-}_k - \mathcal{Q}^{+}_k$, where
\begin{equation}
\mathcal{Q}_k^{\pm} = - 2 k \int_0^1\!\! d\xi \ K^{a_1}_{\pm} \wedge \mathfrak{F}_{0\pm}^{a_2 a_3}(\xi) \wedge \cdots \wedge\mathfrak{F}_{0\pm}^{a_{2k-2} a_{2k-1}}(\xi) \wedge \mathfrak{E}^{(d)}_{a_{1} \cdots a_{2k-1}} ~,
\label{twosideaction}
\end{equation}
where, now, $\mathfrak{F}_{0\pm}^{ab}(\xi) \equiv R_0^{ab} - \xi^2\, K_{\pm}^{a} \wedge K_{\pm}^{b}$. Notice that either term, $\mathcal{Q}^{\pm}_k$, reads like the one discussed in (\ref{LLboundaryK}), and they correspond to each boundary region.  That is, the junction acts as a two-sided boundary. All in all, the $k^{\text{th}}$ contribution to the total action can be written in terms of quantities coming from both sides,
\begin{equation}
\mathcal{I}_k = \left(\int_\mathcal{M_-}\!\!\!\mathcal{L}_k^{-} - \int_\Sigma \mathcal{Q}_k^{-}\right) + \left(\int_\mathcal{M_+}\!\!\!\mathcal{L}_k^{+}+\int_\Sigma\mathcal{Q}_k^{+}-\int_{\partial \mathcal{M}}\!\!\!\mathcal{Q}_k^{+}\right) ~,
\label{splitaction}
\end{equation}
the relative plus sign in the second term coming from the reverse orientation of the surface in that region. The last term is irrelevant for the purposes of the present paper and will not be considered in the following, it just makes the whole outer contribution to vanish if we take the junction surface to infinity. The infinitesimal variation of the boundary action with respect to $\omega_{\pm}$ gives two terms. One is a total derivative. The other cancels with the total derivative term coming from the bulk.

The surface contribution to the equations of motion is just given by the variation with respect to the frame, that correspond on each side to the canonical momentum at the surface. The junction conditions amount just to continuity of the momenta accross the hypersurface $\Sigma$,
\begin{equation}
\pi^+_{ab} = \pi^-_{ab} ~.
\label{junction}
\end{equation}
In the particular case we are interested in, given that all the forms involved in the above expression are diagonal, we should only care about those components.

\subsection{Thermalon configuration}

\begin{figure}[ht]
\centering
\includegraphics[width=.45\textwidth]{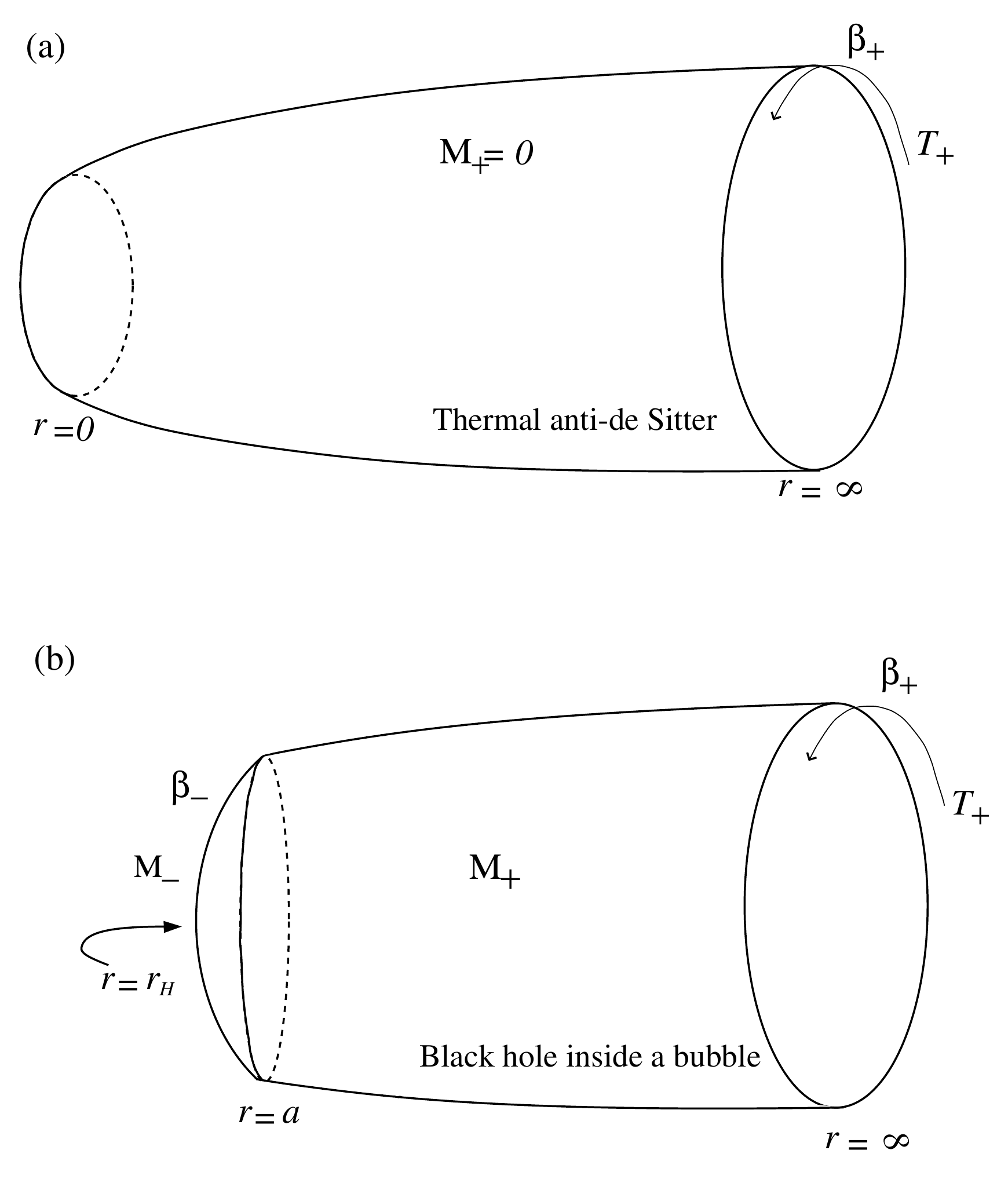}
\caption{Euclidean section of a {\it thermalon} configuration; a {\it bubble} pops out in empty AdS space with cosmological constant $\Lambda_+$, hosting a black hole belonging to a different branch of the Lovelock theory characterized by $\Lambda_-$.}
\label{thermalon}
\end{figure}

Let us be more detailed in the kind of configurations we are interested in. They correspond to a {\it bubble}, whose outer region asymptotes AdS with a cosmological constant $\Lambda_{+}$, while the inner region corresponds to another branch solution characterized by the effective cosmological constant $\Lambda_-$ (see Figure \ref{thermalon}).
The junction conditions will constrain the possible combinations of parameters that will describe the solutions of interest, instantons and thermalons. Across the junction, the vielbein has to be continuous. We thus consider an Euclidean section of the form
\begin{equation}
ds^2 = f_{\pm}(r) dt^2 + \frac{dr^2}{f_{\pm}(r)} + r^2\ d\Omega_{\sigma,d-2}^2 ~,
\label{Eucsection}
\end{equation}
where the $\pm$ signs denote the outer/inner regions. The junction is conveniently described by the parametric equations
\begin{equation}
r = a(\tau) ~, \qquad t_\pm = T_{\pm}(\tau) ~,
\label{parequations}
\end{equation}
with an induced metric of the form
\begin{equation}
ds^2 = d\tau^2 + a(\tau)^2\ d\Sigma_{d-2,\sigma}^{2} ~,
\label{indmetric}
\end{equation}
where we have made the choice
\begin{equation}
f_{\pm}(a)\,\dot{T}^2_{\pm}+ \frac{\dot{a}^2}{f_{\pm}(a)} = 1 ~, \qquad \forall \tau ~.
\label{RTcond}
\end{equation}
The function $a(\tau)$ appears in the induced metric and so it has to be the same seen from both sides, as long as the induced metric itself has to be continuous. The thermalon is characterized by a static configuration, $\dot{a} = \ddot{a} = 0$ ({\it i.e.}, $a(\tau) = a_\star$, the location of the thermalon), that, in view of (\ref{RTcond}), translates into
\begin{equation}
\sqrt{f_{-}(a_\star)}\ T_{-} = \sqrt{f_{+}(a_\star)}\ T_{+} = \tau ~,
\label{Tglue}
\end{equation}
which means that the physical length of the circle in Euclidean time is the same as seen from both sides. This matching condition will eventually allow us to determine the temperature of the solution. In fact, regularity of the Euclidean section (see Figure \ref{thermalon}) demands the inner solution to be a black hole which has an associated Hawking temperature. Thereby, once the periodicity in Euclidean time is fixed to attain a regular horizon, the periodicity in the outer time will be determined through
\begin{equation}
\sqrt{f_{-}(a_\star)}\ \beta_{-} = \sqrt{f_{+}(a_\star)}\ \beta_{+}\equiv \beta_0 ~,
\label{matchbetas}
\end{equation}
where $\beta_{-}$ is the usual inverse Hawking temperature of the inner solution while $\beta_{+}$ is the one seen by an observer at infinity, that will be different from the Hawking temperature corresponding to black holes of the given mass on either branch. $\beta_0$ corresponds to the periodicity in the new variable $\tau$.

Consider the induced vielbein basis $e^{\tau} = d\tau$ and $e^{i} = a(\tau)\ \tilde{e}^{\varphi_i}$, which is intrinsic to $\Sigma$. If we call $x^\mu_\pm(\zeta^a)$ the coordinates of the embedding of $\Sigma$ with intrinsic coordinates $\zeta^a = (\tau,\varphi_i)$ on $\mathcal{M}_\pm$, the unit normal vectors read, in Lorentzian signature,
\begin{equation}
n_\pm^\mu = (\frac{\dot{a}}{f_\pm(a)}, \sqrt{\dot{a}^2 + f_\pm(a)}, 0, \ldots, 0) ~.
\label{normalvectors}
\end{equation}
The extrinsic curvature is then given by
\begin{equation}
K_{\pm\,ab} = -n_{\pm\mu} \left(\frac{\partial^2 x_\pm^\mu}{\partial\zeta^a\partial\zeta^b}+\Gamma^{\mu}_{\ \alpha\beta}\, \frac{\partial x_\pm^\alpha}{\partial\zeta^a}\, \frac{\partial x_\pm^\beta}{\partial\zeta^b}\right) ~,
\label{Kab}
\end{equation}
yielding
\begin{equation}
{K_\pm}^\tau_{\ \tau} = \frac{\ddot{a} + \frac12 f'_\pm(a)}{\sqrt{\dot{a}^2+f_\pm(a)}} ~, \qquad {K_\pm}^{\varphi_i}_{\ \varphi_j} = \frac{\sqrt{\dot{a}^2+f_\pm(a)}}{a}\ \delta^i_{\ j} ~.
\end{equation}
The intrinsic curvature, in turn, is given by
\begin{equation}
R^{i\tau}_0=\frac{\ddot{a}}{a}\,e^{i}\wedge e^{\tau} ~, \qquad R^{ij}_0 = \frac{\sigma+\dot{a}^2}{a^2}\,e^{i}\wedge e^{j} ~.
\label{R0ab}
\end{equation}
Notice that it is trivially the same as seen from either side as it is calculated from the induced metric and this is continuous. The vielbein basis is continuous but the spin connection is not. It is clear that all the components aligned along the normal direction of this intrinsic spin connection are zero in the same way as it happens for the corresponding intrinsic curvature. In Euclidean signature we just change the signs of the squared velocity and acceleration as $(\dot{a}^2,\ddot{a})\ \to \ (-\dot{a}^2_E,-\ddot{a}_E)$.

\subsection{Bubble dynamics}

The junction conditions (\ref{junction}) for the configurations of interest have just diagonal components related by a conservation equation (Bianchi identity) that constrains them in such a way that only the $\tau\tau$ component matters. The remaining components are related \cite{Davis2003,wormholes}
\begin{equation}
\frac{d}{d\tau}\left(a^{d-2}\, \pi^\pm_{\tau\tau}\right) = (d-2)\, a^2 \dot{a}\, \pi^\pm_{\varphi_i\varphi_i} ~, \qquad \forall i ~,
\label{Bianchi}
\end{equation}
such that if $\pi^\pm_{\tau\tau}$ verifies (\ref{junction}), all the components automatically do. Thus, we just need to compute $\Pi^\pm \equiv \pi^\pm_{\tau\tau}$ (we introduce $\Pi^\pm$ to avoid the use of indices). This involves just the {\it angular} components of both the intrinsic and extrinsic curvatures; the expression reduces to
\begin{equation}
\Pi^\pm = \frac{\sqrt{\dot{a}^2+f_\pm(a)}}{a} \int_0^1\! d\xi\ \Upsilon'\left[\frac{\sigma-\xi^2\,f_\pm(a)+ (1 - \xi^2)\,\dot{a}^2}{a^2}\right] ~,
\end{equation}
where we avoid the inclusion of some factors that are irrelevant in our discussion since we are not considering the inclusion of matter and, thus, ultimately will impose $\Pi^+ = \Pi^-$. Notice that the polynomial $\Upsilon$ is again seen to play a central r\^ole.

If we define $\widetilde{\Pi} \equiv \Pi^+ - \Pi^-$, the junction condition (\ref{junction}) and Bianchi identity (\ref{Bianchi}) can be written as $\widetilde{\Pi} = \frac{d\widetilde{\Pi}}{d\tau} = 0$. Notice that the space-time dimensionality is somehow irrelevant in this expression. In the case of LGB, it reduces to the one in \cite{Garraffo2008b}. If we conveniently introduce
\begin{equation}
g_\pm \equiv g_\pm(a) = \frac{\sigma-f_{\pm}(a)}{a^2} ~, \qquad {\rm and} \qquad H \equiv H(a,\dot{a}) = \frac{\sigma + \dot{a}^2}{a^2} ~,
\end{equation}
we can rewrite
\begin{equation}
\Pi^\pm\,[g_\pm, H] = \sqrt{H-g_\pm}\int_0^1\! d\xi\ \Upsilon'\left[\xi^2\,g_\pm + (1 - \xi^2)\,H \right] ~,
\label{Sjunc}
\end{equation}
where it becomes clear that all the information about the branches is contained in $g_\pm$. Notice that $g_\pm \leq H$ in order to have a real value for $\Pi^\pm$ (which tantamounts to $f_\pm(a) \geq 0$ in order to have a real Euclidean boundary action). This implies that the static bubble corresponding to the {thermalon} necessarily forms at $a_\star \geq \max(r_{H+},r_{H-})$, outside the would be black hole horizons corresponding to both branches. Moreover, this reality condition is also necessary for the equation to yield real values of the velocity.

The difference in canonical momenta $\widetilde{\Pi}$ may be rewritten as
\begin{equation}
\widetilde{\Pi} = \left(\sqrt{H-g_+}-\sqrt{H-g_-}\right) \int^1_0\! d\xi\ \Upsilon'\left[H-\left(\xi\sqrt{H-g_+}+(1-\xi)\sqrt{H-g_-}\right)^2\right] ~,
\label{totaljunc}
\end{equation}
which is the variation of (\ref{matchactionbis}). A convenient change of integration variable results in
\begin{equation}
\widetilde{\Pi} = \int_{\sqrt{H-g_-}}^{\sqrt{H-g_+}}\!dx\ \Upsilon'[H-x^2] ~,
\label{changePi}
\end{equation}
which is more compact and easier to manipulate. Recall that $\widetilde{\Pi} = \widetilde{\Pi}(\dot{a},a)$. We can now interpret $\widetilde{\Pi} = 0$ as a conservation equation with a non-canonical kinetic term. We can make this statement more precise by noticing that
\begin{equation}
\Pi_+^2 = \Pi_-^2 \qquad \Longleftrightarrow \qquad \prod_{i=1}^{K-1}\left(\frac12\dot{a}^2+V_i(a)\right) = 0 ~,
\label{newt}
\end{equation}
where we have to take into account that some of the {\it roots} or potentials, $V_i(a)$, correspond to the actual junction conditions (\ref{junction}), while some others are associated to the reversal orientation, $\Pi^+ = - \Pi^-$, that amounts to gluing two interiors or two exteriors (a {\it wormhole} configuration). Besides, some roots shall be discarded in case they yield imaginary momenta, $g_\pm > H$. This reality condition amounts to
\begin{equation}
\dot{a}^2 \geq -f_{\pm} ~, \qquad {\rm or} \qquad \dot{a}^2_E \leq f_\pm ~,
\end{equation}
in the Lorentzian and Euclidean sections respectively. Notice that in the Euclidean version the bubble is bound to propagate outside the event horizon whereas in the Lorentzian case it may cross it. This might lead to a seemingly pathological situation, namely, the destruction of the horizon by a collapsing bubble ultimately leading to a violation of the cosmic censorship hypothesis. Even though we expect this possibility to be ruled out by the kind of mechanisms studied in \cite{CamanhoE4}, some comments are in order.

Roots of $\widetilde{\Pi} = 0$ and $\Pi^+ + \Pi^- = 0$ may just join where $\Pi^+ = \Pi^- = 0$. For instance, whenever $H = g_-$ or, equivalently, $\dot{a}^2 = -f_-$, something that, as (\ref{newt}) makes manifest, is possible only inside the horizon or for negative values of $\dot{a}^2$, a forbidden region of the potential. In fact, in these particular points the potential ceases to be a solution of $\Pi^+ - \Pi^- = 0$ to become a solution of $\Pi^+ + \Pi^- = 0$. This is a generic feature of any Lovelock theory since $\sqrt{H-g_-}$ does not change sign at $H=g_-$, whereas $\Pi^-$ does. When this happens inside the horizon it is unclear what happens to the bubble. It cannot go further but it cannot turn back as it would be traveling backwards in time. The potential becomes unphysical beyond the point where it meets $f_-/2$, which corresponds to $\dot{a}^2 = -f_-$. Notice that we could in principle get to the origin if we reduce $M_+$ or, for fixed $M_+$ if we increase $M_-$, actually for $M_+\leq 4M_-$.
\begin{figure}[ht]
\centering
\includegraphics[width=.58\textwidth]{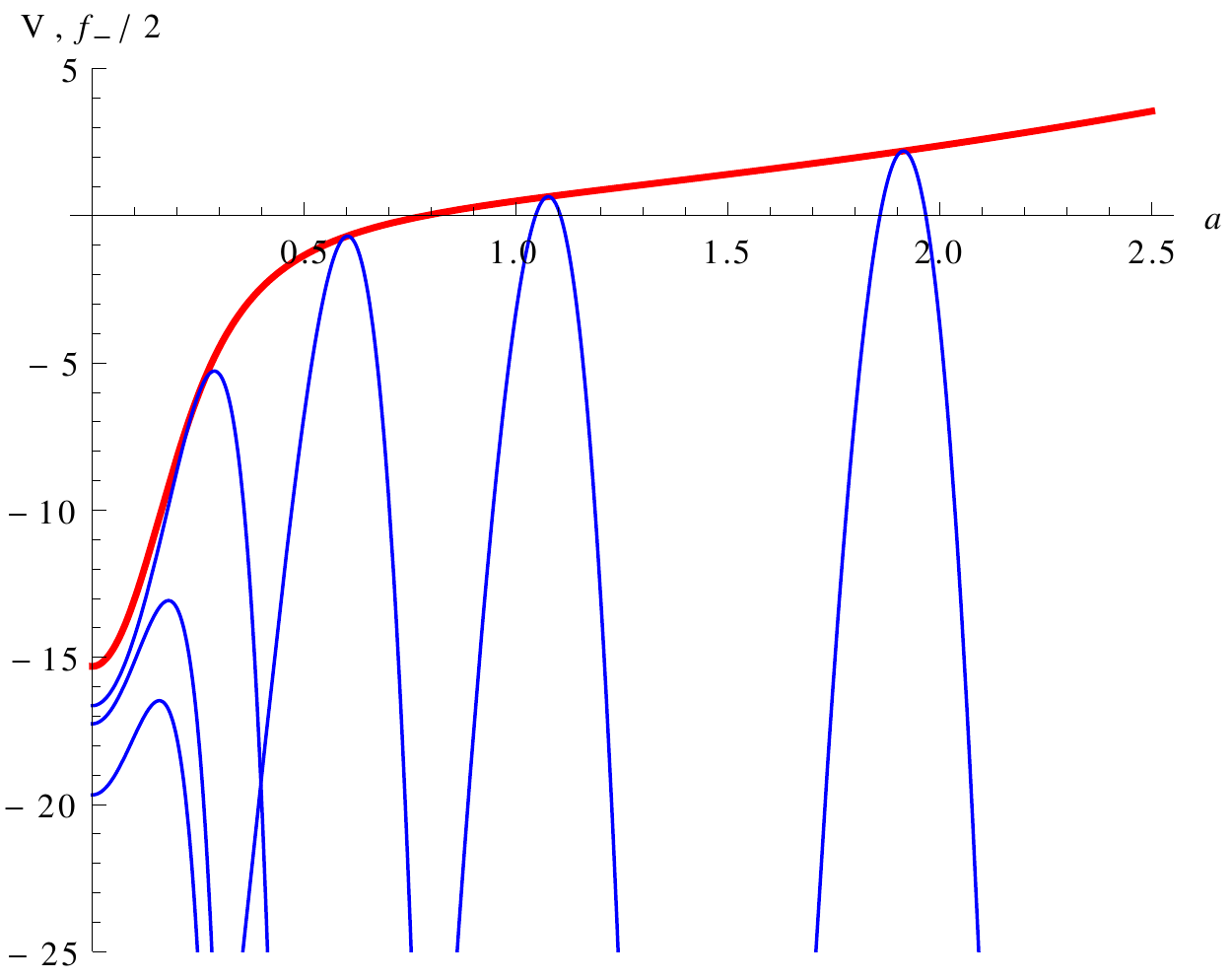}
\caption{Potential (in blue, thin) of a brane gluing spherically symmetric stable and unstable branches of LGB gravity with parameters $\lambda=0.001$, $L=1$, $M_-= 1$ and $M_+=2,2.4,7,10^2,10^3,10^4$ from left to right. $f_-/2$ (in red, thick) is (half the) metric function corresponding to the stable branch. The blue curve slides up the red one as we increase $M_+$. For the first two values of $M_+$ the blue and red curves do not intersect.}
\label{rootchange}
\end{figure}
Figure \ref{rootchange} displays these facts in the simplest case of the LGB theory. We will examine further aspects of the collapsing bubble scenario in section \ref{collapse}.

Once a particular potential is chosen, say $V_j(a)$, we can use (\ref{Bianchi}) to determine the corresponding acceleration equation that governs the dynamics of the bubble
\begin{equation}
\ddot{a} = -V_j'(a) ~.
\label{dynbubble}
\end{equation}
Notice that this dynamics may be difficult to determine, in general, since for generic $K$ it might be impossible to have an explicit expression for the potentials, and, on top of that, several of them may provide a suitable dynamics for the bubble. In all the expressions the r\^oles of the two branches can be exchanged yielding the same bubble dynamics. As there is no matter on it, the corresponding equations are blind to which is the inner/outer solution, the bubble behaving in exactly the same manner.

\subsubsection{Thermalons: stability and existence}

There are two limiting cases of the junction conditions that are of special interest. On the one hand we are interested in the static configurations ({thermalons}) and their stability. For that it is enough to consider the slow limit of (\ref{junction}), in a double expansion about $a_\star$ and $\dot{a}=0$, 
\begin{equation}
\widetilde{\Pi}\approx \widetilde{\Pi}^\star + \frac{\partial \widetilde{\Pi}^{\,\star}}{\partial H}\, \frac{\dot{a}^2}{a_\star^2} + \frac{\partial \widetilde{\Pi}^{\,\star}}{\partial a}\, (a-a_\star)+\frac12 \frac{\partial^2 \widetilde{\Pi}^{\,\star}}{\partial a^2}\, (a-a_\star)^2 ~,
\end{equation}
where the upper star means that a quantity is being evaluated after the replacement $a \to a_\star$ and $\dot{a} \to 0$, {\it e.g.}, $H \to H_\star \equiv \frac{\sigma}{a_\star^2}$ and $g_\pm \to g_\pm^\star \equiv g_\pm(a_\star)$. Besides the two conditions
\begin{equation}
\widetilde{\Pi}^\star = \frac{\partial \widetilde{\Pi}^\star}{\partial a} = 0 ~,
\label{junctionsc}
\end{equation}
the junction rule (\ref{junction}) at $a = a_\star$ adopts the canonical form of an energy constraint for an auxiliary harmonic system described by $a(\tau)$,
\begin{equation}
\frac12 \dot{a}^2 + \frac12 k\, (a-a_\star)^2 = 0 ~, \qquad {\rm with} \quad k = \frac{a_\star^2}{2} \left( \frac{\partial \widetilde{\Pi}^{\,\star}}{\partial H} \right)^{-1}\! \frac{\partial^2 \widetilde{\Pi}^{\,\star}}{\partial a^2} ~,
\label{potstability}
\end{equation}
provided the denominator is non-vanishing, which tantamounts to requiring the potential to be a smooth function of the radius at $a_\star$. The factor vanishes when two potentials merge, thus becoming complex beyond that point. The bubble cannot go inside the region of complex potential; analogously to the case of branch singularities in the cosmological context, it will turn back along another {\it root} that merges with the one it was following \cite{XianPhD}.

\begin{figure}[h]
\centering
\includegraphics[width=.58\textwidth]{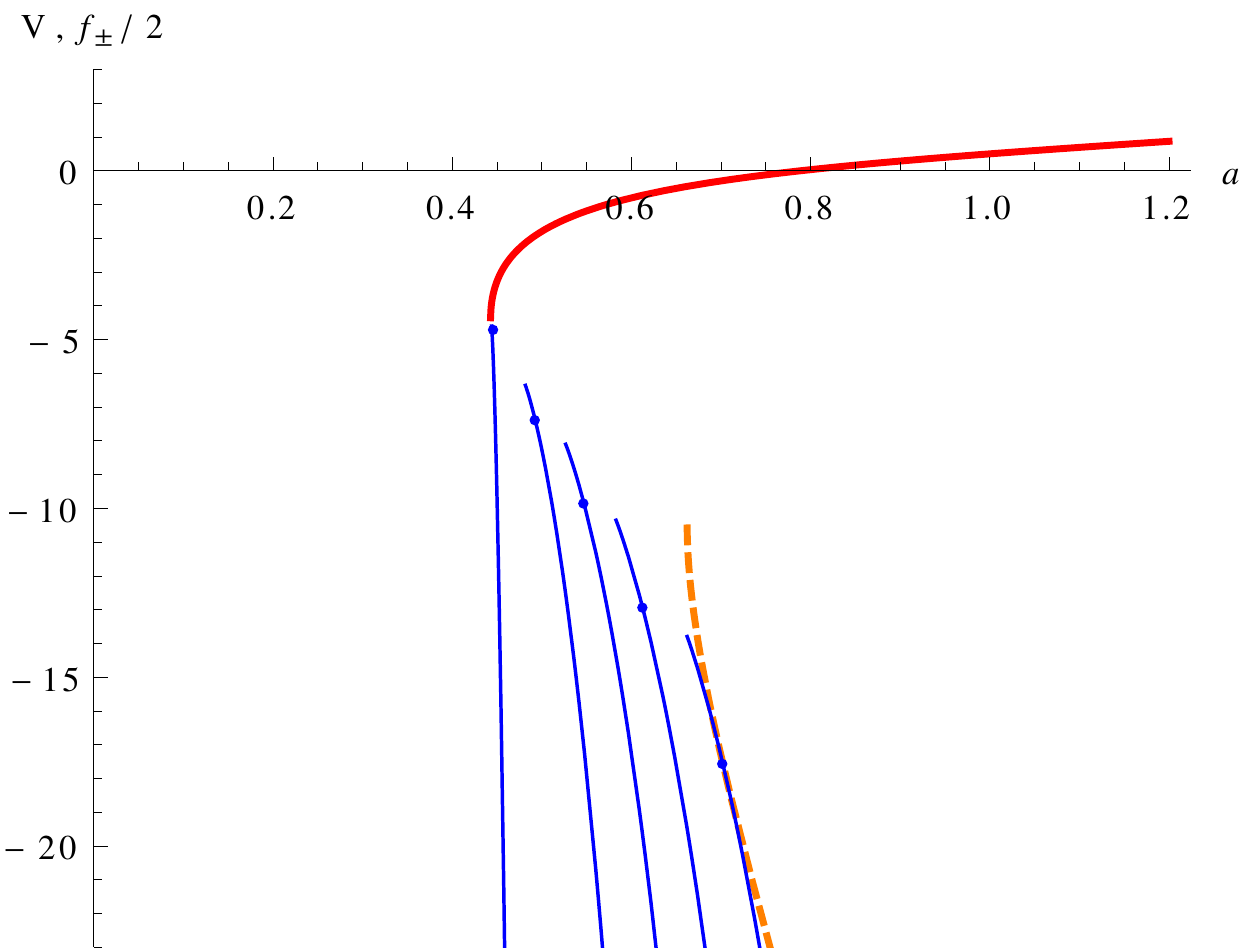}
\caption{Potential (in blue, thin) of a brane gluing spherically symmetric stable and unstable branches of LGB gravity with parameters $\lambda=-0.01$, $L=1$, $M_-=1$ and $M_+= 1.01,1.4,2,3,5$ from left to right. $f_-/2$ (in red, thick) is (half the) metric function corresponding to the stable branch and $f_+/2$ (in orange, dashed) is (half) the one corresponding to the unstable branch (for $M_+=5$). Singularities are located where the curves end, the potential ending when we find the outermost of them. Before that we encounter a point where $V=f_+/2$ (dots) beyond which the potential is unphysical. The physical potential does not even display a maximum for the thermalon to sit.}
\label{braneHorizon}
\end{figure}

It is important to stress on the fact that the existence of the thermalon should not be taken for granted. An obvious situation in which there is no thermalon configuration is that with an inner unstable branch: its Euclidean section being singular. But a more subtle case is nicely exemplified by the LGB theory with negative $\lambda$ coupling (see Figure \ref{braneHorizon}).
There is a value of $a$ where $V=f_+/2$ (the dots in the Figure) beyond which the potential is unphysical. This removes the would be top of the hill (which in the Euclidean section becomes a local minimum) sweeping off any possible thermalon configuration. 

\subsubsection{Fate of the bubbles}
\label{fatebubble}

The other interesting limiting case corresponds to the speed of the bubble becoming large since it will be relevant when discussing its asymptotic run away behavior. As $a\rightarrow\infty$, the behavior of $H$ has to be given by a power law, and so it can either diverge or asymptote to a constant. We can verify that one of the solutions diverges as $H\sim a^{d-1}$ whereas the remaining ($K-2$) are asymptotically constant. We can prove that the maximum number of solutions is in general $K-1$ just by squaring each side of the $\Pi^+ = \Pi^-$ equation in order to obtain a polynomial equation on $H$. Even if the na\"ive degree of such equation is $2K-1$, it is easy to show that the first non-vanishing coefficient is order $K-1$ in $H$.

In the limit of large bubble speed, equation (\ref{totaljunc}) can be evaluated directly using
\begin{eqnarray*}
H-\left(\xi\sqrt{H-g_+}+(1-\xi)\sqrt{H-g_-}\right)^2&\approx & \xi g_+ + (1-\xi)g_-+\frac{1}{4H}\xi(1-\xi)\left(g_+ - g_-\right)^2 ~, \\ [0.58em]
\left(\sqrt{H-g_+}-\sqrt{H-g_-}\right)&\approx &\frac{1}{2\sqrt{H}}\left(g_- -g_+\right)\left(1+\frac1{4H}(g_-+g_+)\right) ~,
\end{eqnarray*}
so that we can verify, to next-to-leading order in $1/H$, that 
\begin{equation}
\widetilde{\Pi} \approx -\frac{1}{2\sqrt{H}}\left[ \Upsilon[g_+]-\Upsilon[g_-] - \frac{1}{2H}\left(\int_{g_-}^{g_+} \!dx\ \Upsilon[x]-\left(g_+\Upsilon[g_+]-g_-\Upsilon[g_-]\right)\right)\right] ~.
\label{inview}
\end{equation}
Here we used the change of variable $x = \xi g_+ + (1-\xi) g_-$, thus solving for
\begin{equation}
H\approx\frac{1}{2(M_+ - M_-)}\left[a^{d-1}\int_{g_-}^{g_+} \!dx\ \Upsilon[x]-(g_+ M_+ - g_- M_-)\right] ~.
\label{highH}
\end{equation}
In the lightlike limit ($H\rightarrow\infty$),  we get a consistent equation for $H$ either if $a\rightarrow\infty$ and $H\sim a^{d-1}$ or $M_+ \to M_-$, which coincides with the result of \cite{Gravanis2009}. In the former case, the asymptotic behavior of the potential reads
\begin{equation}
H \approx \frac{a^{d-1}}{2(M_+ - M_-)} \int_{\Lambda_-}^{\Lambda_+} \!dx\ \Upsilon[x] ~,
\label{asymptH}
\end{equation}
and this solution is {\it physical} in the sense of being a solution of $\widetilde{\Pi}=0$ (as opposed to a $\Pi^+ + \Pi^- = 0$ root). The asymptotic solution (\ref{asymptH}) is useful to analyze in which cases it is possible for the bubble to reach infinity. For that, one also has to analyze the sign of the asymptotically constant solutions, when present, and also determine along which branch the bubble is propagating. The constant roots are the solutions of 
\begin{equation}
\widetilde{\Pi}[\Lambda_\pm,H] = 0 ~,
\end{equation}
that is degree $K-2$ in $H$, given that, in view of (\ref{inview}), the leading power is proportional to $\Upsilon[g_+]-\Upsilon[g_-]\sim 1/a^{d-1}$, and it vanishes when we approach the asymptotic region. In that limit, the topology of the horizon (given by $\sigma$) becomes irrelevant.

In case the bubble may run away to infinity, it asymptotically approaches the speed of light. This can be seen for instance from (\ref{asymptH}). Since $\dot{a}^2$ grows faster than the function $f$ then the radial speed, 
\begin{equation}
\frac{dr}{dt} = \frac{\dot{a}}{\sqrt{f+\dot{a}^2}}\; f \rightarrow f ~,
\end{equation}
is exactly the same as for a null geodesic in that background. In that way, as AdS space has a timelike boundary, the bubble gets there in a finite time and, as such, it can be interpreted as a change of boundary conditions for the theory, {\em i.e.}, a jump from one branch of solutions to another. The time it takes for the bubble to go from any position, $a_0$, to the boundary is actually
\begin{equation}
\Delta \tau = \int_{a_0}^\infty \frac{da}{\dot{a}}\sim \int_{a_0}^\infty \frac{da}{a^{\frac{d+1}{2}}} \sim  a_0^{-\frac{d-1}{2}} < \infty ~;
\end{equation}
and it is also finite for the asymptotically constant values of $H$, although the asymptotic speed becomes a fraction of the speed of light in that case.

In the case of LGB gravity there is just one possible potential that determines the dynamics of the bubble separating solutions belonging to the two different branches. It can be simply written as (recall that $g_\pm = g_\pm(a)$)
\begin{equation}
V(a) = \frac{a^{d+1}}{24\lambda (M_+ - M_-)} \left[ g_+ (3 + 2\lambda g_+)^2 - g_- (3 + 2\lambda g_-)^2 \right] +\frac{\sigma}{2} ~.
\label{GBpots1}
\end{equation}
It verifies all of the general properties described above. For instance, as the integral of $\Upsilon[g]$ between $\Lambda_-$ and $\Lambda_+$ is always positive, the bubble can escape to infinity as long as the mass of the unstable branch is larger than that of the stable one, $M_+ > M_-$. Remember that any of the two can be in principle on either side of the bubble and the dynamics is exactly the same. The existence of a thermalon configuration disfavor this {\it a priori} symmetric setup. The tension depends on the position and there is no pressure term as it would depend on the side corresponding to each branch. The bubbles have tension despite of the fact that they contain no matter.

\subsubsection{Collapsing bubbles and cosmic censorship}
\label{collapse}

An expanding bubble looks certainly more physical, at least whenever the unstable branch is outside. In that case, one may think of a bubble popping out from a naked singularity (or from thermal AdS space) and creating a horizon, rapidly expanding, and leaving a regular black hole behind. This points towards a possible instability that triggers a phase transition through the formation of bubbles of the new phase. This phenomenon is the main target of the present paper and will be explored in detail starting in the following section. But we would like to discuss first a few aspects of the {\it a priori} less interesting but also a bit puzzling situation of the equally probable collapsing bubbles.

Depending upon the values of the masses, the stable branch of the LGB theory may have an event horizon and the naked singularity of the unstable solution may be located at either side. Quite generally ({\em e.g.}, see Figure \ref{rootchange} for low enough $M_+$) the horizon is accessible in the sense that the potential is finite and negative at that radius. Assuming that the stable branch is the inner one, there is nothing to prevent a collapsing bubble from reaching the event horizon. It would necessarily undergo subsequent collapse until it reaches the naked singularity that would then become visible to an external observer. For that we just need the radius of the horizon to be larger than the radius of the branch singularity, otherwise the latter would become naked even before the bubble reaches any horizon. In the opposite case, with the unstable branch inside, the contraction of the brane would instead create a horizon. We will provide specific examples below.

The destruction of the horizon by the bubble might seem unnatural but it seems to be what really happens. In order to verify that this is indeed the case one has to be careful and check that nothing goes wrong at the horizon or inside it. In particular, in order for the metric to be continuous, the existence of a common induced metric on the junction surface is not enough. We also have to ensure that the change of variables between the coordinate frames on both sides is regular. In the vicinity of the junction ($\rho=0$) we might write the metric in terms of  a coordinates set adapted to the surface
\begin{equation}
ds^2 \approx -d\tau^2+d\rho^2 + a^2(\tau)\, d\Sigma_{\sigma,d-2}^2 ~,
\end{equation}
the same in both sides, $\rho$ being the normal. We then get constraints on the coordinate functions of the brane $T_\pm(\tau,\rho)$ and $a_\pm(\tau,\rho)$ (such that $T_\pm(\tau,0)\equiv T_\pm(\tau)$ and $a_\pm(\tau,0)\equiv a(\tau)$): the Lorentzian analog of (\ref{RTcond}),
\begin{equation}
-f_\pm\, {T'}_{\!\!\!\pm}^2+\frac{{a'}_{\!\!\!\pm}^2}{f_\pm} = 1 ~,
\end{equation}
and, additionally,
\begin{equation}
-f_\pm\, {T'}_{\!\!\!\pm}\,\dot{T}_\pm+\frac{{a'}_{\!\!\!\pm}\,\dot{a}_\pm}{f_\pm} = 0 ~,
\end{equation}
where primes indicate derivatives with respect to the new variable $\rho$. The radial variables are not equal in this case even though they  are at the junction. In this way, we may write the change of variables as 
\begin{equation}
dt_\pm = \dot{T}_\pm\, d\tau+\frac{\dot{a}}{f_\pm}d\rho ~, \qquad dr_\pm = \dot{a}\, d\tau + f_\pm\, \dot{T}_\pm\, d\rho ~,
\label{dtaudrho}
\end{equation}
that corresponds to a boost in the $(t_\pm,r_\pm)$-plane. This can be easily verified using the orthonormal frame
\begin{equation}
\left( \begin{array}{c} e^0_\pm \\ e^1_\pm \end{array} \right) = U_\pm \left( \begin{array}{c} d\tau \\ d\rho \end{array} \right) ~, \qquad U_\pm = \frac{1}{\sqrt{f_\pm}} \left( \begin{array}{cc} f_\pm\,\dot{T}_\pm & \dot{a} \\
\dot{a} & f_\pm\,\dot{T}_\pm \end{array} \right) ~,
\end{equation}
the boost matrix having unit determinant while the inverse can be obtained just by changing the sign of $\dot{a}$. Recall that $\dot{T}_\pm$ can be written in terms of $\dot{a}$ by means of (\ref{RTcond}), and it turns out to be invariant under such change. Thus, the change of variables between inner and outer coordinates corresponds to a composition of boosts $e_+=U_+U^{-1}_-e_-$, a transformation that does not change the causal structure, {\em i.e.}, null geodesics are continuous across the bubble.

\begin{figure}[ht]
\centering
\begin{tikzpicture}[scale=0.76]
\node (I)    at ( 4,0)   {};
\node (II)   at (-4,0)   {};
\node (III)  at (0, 2.5) {$(-)$};
\node (IV)   at (0,-2.5) {};
\node (boosted) at (45:8) {};
\path  
  (II) +(90:4)  coordinate (IItop)
       +(-90:4) coordinate (IIbot)
       +(0:4)   coordinate (IIright)
       +(180:0) coordinate (IIleft)
       ;
\draw[thick] (IItop) -- (IIright);
\draw (IIright) -- (IIbot) --  (IItop);
\path 
   (I) +(90:4)  coordinate (Itop)
       +(-90:4) coordinate (Ibot)
       +(180:4) coordinate (Ileft)
       +(0:0)   coordinate (Iright)
       ;	
\path 
	 ($(IItop)!.75!(Itop)$) coordinate (boostBL)
							+(60:2) coordinate (boostTL)
							;
\path
	 (boosted)  +(-120:.2) coordinate (boostBR)
							+(60:2) coordinate (boostTR)
							;
\draw[thick]  (Iright) --
		 node[midway,left] {$(+)$}
		 (Ibot) -- (Ileft) --
		 node[midway, above, sloped] {$r=r_H$}
				($(Ileft)!.5!(Itop)$);
\draw[decorate,decoration=zigzag,thick] (IItop) -- (boostBL) 
      node[midway, above, inner sep=2mm] {$r=0$};
\draw[decorate,decoration=zigzag] (IIbot) -- (Ibot)
			node[midway, below, inner sep=2mm] {$r=0$};	
\draw[dashed,thick] 
    ($(Ileft)!.5!(Iright)$) to[out=90, in=-90, looseness=1.5] (boostBL);
\draw[dashed,thick] 
    (Ibot) to[out=125, in=-90, looseness=1] ($(Ileft)!.5!(Iright)$);
\draw[thick]
    (Iright) to[out=90, in=-120, looseness=1]
		node[midway,left] {$(+)$}
		(boosted) -- (boostTR);		
\draw[decorate,decoration=zigzag,thick] 
    (boostBL) to[out=60, in=-120, looseness=1] (boostTL);		
\draw[dashed] 
    (boostBL) -- (boosted);	
\draw[dashed] 
    (boostTL) -- (boostTR);	
\draw[dashed] 
    (Ileft) --
		node[below] {$(-)$}
		($(Ileft)!.65!(Iright)$) -- 
		node[above] {$t=0$}
		(Iright);
\end{tikzpicture}
\caption{Schematic Penrose diagram for a {\it nearly} static bubble that undergoes collapse from $t=0$ onwards. The thick dashed line represents the trajectory of the bubble with inner black hole and outer naked singularity geometries. The spacelike singularity becomes timelike as the bubble reaches $r=0$. The resulting space-time is a (boosted) naked singularity, dashed lines representing constant $t_+$ slices. The lowest one corresponds to the Cauchy horizon introduced by the timelike singularity. The upper and right wedges represent the physical region of the space-time corresponding to a black hole formed by collapse whereas the rest corresponds to the complementary white hole region.}
\label{collPenrose}
\end{figure}
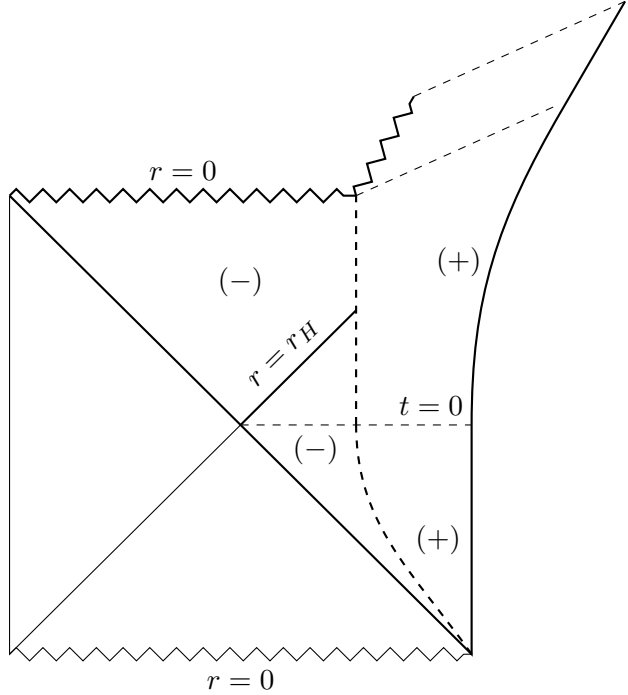

We can now address the question of the behavior of the bubble as we cross the horizon. In that case we have to change $\sqrt{f_-}=i\sqrt{|f_-|}$, as $f_-$ goes negative, in such a way that the timelike and spacelike vielbein exchange their r\^oles\footnote{Indeed, notice that, inside the horizon, $\hat{e}^0 = -\frac{dr}{\sqrt{|f|}}$ and $\hat{e}^1 = \sqrt{|f|}\,dt$.} preserving the orientation ($e^0_-\wedge e^1_-=\hat{e}_-^0\wedge \hat{e}_-^1$),
\begin{equation}
e^0_- = i\hat{e}^1_- ~, \qquad e^1_- = i\hat{e}_-^0 ~.
\end{equation}
Consequently, the change of variables behind the horizon reads
\begin{equation}
\left( \begin{array}{c} \hat{e}^0_- \\ \hat{e}^1_- \end{array} \right) = \hat{U}_- \left( \begin{array}{c} d\tau \\
d\rho \end{array} \right) ~, \qquad \hat{U}_- = - \frac{1}{\sqrt{|f_-|}} \left( \begin{array}{cc} \dot{a} & \sqrt{f_-+\dot{a}^2} \\ \sqrt{f_-+\dot{a}^2} & \dot{a} \end{array} \right) ~,
\label{changebeyond}
\end{equation}
which again corresponds to a boost and is completely well defined. The change is actually continuous across the horizon, the bubble being able to cross it. After all, there is nothing special\footnote{This assertion may need to be revised or refined under the light of the recently proposed {\it firewall paradigm} \cite{AMPS}.} --locally-- about that point. Besides, once it gets to the horizon the causal structure makes it impossible for the bubble to get back (see Figure \ref{collPenrose} for the Penrose diagram of the process) as it would be traveling backwards in time. 
This can also be seen from (\ref{changebeyond}), as long as the diagonal components of the $\hat{U}_-$ matrix have to be positive and bigger than one. This is also true for $U_\pm$ outside the horizon.

It is remarkable that, even though the original horizon is destroyed by the bubble the inherited causal structure makes it impossible for it to go back. In this case we cannot blame the symmetry of the solutions considered. Once the horizon is crossed, its (previous) presence forces the bubble to actually reach the central singularity, regardless of the details of the collapse. The singularity unavoidably becomes naked, this being a clear violation of the cosmic censorship hypothesis. Nonetheless, this violation might be a marginal one if, due to the instabilities suffered by those singular solutions, the energy leaks the singularity, thus leaving regular pure AdS$_+$ behind, a process that is allowed by the causal structure. Notice that the bubble itself may emit part of its mass, $M_+-M_-$, to infinity before it actually reaches the singularity. However the naked singularity will still form as the mass contained in the inner solution, $M_-$, cannot escape until the bubble reaches the center.

\section{Thermalons and thermodynamics}

Static bubble configurations are subject to the same discussion that concerns the thermodynamic aspects of black holes. In very much the same way, the new solutions will be characterized by the same thermodynamic variables. In this Section we will compute the value of the Euclidean on-shell action for these {thermalons}. Such static solutions will exist in some cases as can be seen in Figure \ref{rootchange} for the case of LGB gravity with positive $\lambda$ coupling. Being static, they trivially have a smooth Euclidean section whenever they display a horizon in the inner solution covering the singularity. We have to choose the periodicity in Euclidean time accordingly, so that the full configuration is smooth. Once the inner periodicity is fixed to attain a regular horizon, the periodicity in the outer time will be determined through (\ref{matchbetas}), thus fixing the temperature.

As for black holes, the Euclidean on-shell action is in general divergent and needs to be regularized, either by background subtraction or by other means. We will measure the free energy with respect to the specific solution used as a groundstate, usually the maximally symmetric one. In order to simplify the discussion we will calculate the on-shell action in terms of three parameters: the equilibrium position of the bubble, $a_\star$, and the inner/outer periodicity in Euclidean time, $\beta_\pm$. It is important to keep in mind that these two variables are not independent from each other. We will use $(+)$ to denote the outer region and $(-)$ for the inner one. This is (more general but) consistent with the notation of the LGB case explored in \cite{Camanho2012}.

Unlike the computation of the Hawking-Page transition where the fields are continuous, here we have to consider the contribution to the action arising from the bubble, when writing the Euclidean action, (\ref{splitaction}), in the form 
\begin{equation}
\widehat{\mathcal{I}} = \widehat{\mathcal{I}}_- + \widehat{\mathcal{I}}_{\Sigma} + \widehat{\mathcal{I}}_+ ~.
\label{fullaction}
\end{equation}
The outer piece, $\widehat{\mathcal{I}}_+$, includes all the boundary terms at infinity necessary to have a well defined variational principle. We regularize its divergence by subtracting the background $M_+ = 0$ with the same periodicity at infinity, yielding\footnote{In the following expressions, we are constrained to use for the Euclidean action the same normalization utilized for the mass in (\ref{eqg}); {\it i.e.}, we omit a factor $((d-2)!\,V_{d-2})^{-1}$.}
\begin{equation}
\widehat{\mathcal{I}}_+(a_\star,\beta_+)= \beta_+ \left. \partial_r\left[r^d\,\widetilde\Upsilon[g_+]\right]\right|_{a_\star} ~.
\label{Iout}
\end{equation}
It is the same expression as the Euclidean action of the usual black holes, but evaluated at the position of the bubble instead of the horizon. In the precedent formula, we introduced the polynomial \cite{CamanhoE3}
\begin{equation}
\widetilde\Upsilon[g]=\sum_{k=0}^K{\frac{c_k}{d-2k}\,g^k}~,
\end{equation}
related to the characteristic polynomial by means of
\begin{equation}
\Upsilon[g] = d\widetilde\Upsilon[g] - 2g \widetilde\Upsilon'[g] ~, \qquad \Upsilon'[g] = (d-2) \widetilde\Upsilon'[g] - 2g\widetilde\Upsilon''[g] ~.
\label{twopoly}
\end{equation}
The term $\widehat{\mathcal{I}}_-$, in turn, is integrated from the horizon to the location of the bubble yielding two terms of the same form, evaluated at the bubble and the horizon radii 
\begin{equation}
\widehat{\mathcal{I}}_-(a_\star,\beta_-) = \beta_- \left(\left. \partial_r\left[r^d\,\widetilde\Upsilon[g_-]\right]\right|_{r_H} - \left. \partial_r\left[r^d\,\widetilde\Upsilon[g_-]\right]\right|_{a_\star} \right) ~,
\label{Iin}
\end{equation}
where $r_H$ is the radius of the horizon, if any, or zero. Finally, $\widehat{\mathcal{I}}_\Sigma$ is given by
\begin{equation}
\widehat{\mathcal{I}}_\Sigma = \widehat{\mathcal{I}}_\partial^+(a_\star,\beta_0) - \widehat{\mathcal{I}}_\partial^-(a_\star,\beta_0) ~,
\end{equation}
where the periodicity in Euclidean time is inherited from the bulk regions, $\beta_0=\sqrt{f_\pm(a)}\beta_\pm$. Then, we can collect all the contributions that depend upon the location of the bubble,
\begin{eqnarray}
\widehat{\mathcal{I}}_{bub}(a_\star,\beta_0) & = & \beta_0 \left(\frac1{\sqrt{f}} \partial_r\left.\left[r^d\, \tilde\Upsilon[g]\right]\right|_{a_\star}\right.\nonumber\\
& & \qquad + 2\left.\left.\partial_r\left.\left[ r^{d-2} \sqrt{f}\;\int_0^1 \!d\xi\ \tilde\Upsilon'\left[(1-\xi^2)g+\xi^2\, g\right]\right]\right|_{a_\star}\right)\right|_-^+ ~,
\label{Ibrane2}
\end{eqnarray}
in the static case, otherwise we cannot perform the $\tau$ integration, and where $\left.\mathcal{F}(g)\right|^+_-$ is used to indicate the difference between the terms evaluated on both sides of the bubble, $\left.\mathcal{F}(g)\right|^+_- = \mathcal{F}(g_+) - \mathcal{F}(g_-)$. 
Whatever remains shall be consequently called $\widehat{\mathcal{I}}_{bh}$,
\begin{equation}
\widehat{\mathcal{I}}_{bh} = \widehat{\mathcal{I}}-\widehat{\mathcal{I}}_{bub} = \beta_-M_--S~,
\end{equation}
the familiar contribution from the inner black hole, normalized in accordance with our present conventions, yields an entropy, $S = 4\pi r_H^{d-2} \widetilde\Upsilon'[g_H]$. The surface term trivially vanishes when the same solution is taken in both sides of the junction. The contribution from the bubble can be greatly simplified making use of the relations between the two characteristic polynomials (\ref{twopoly}), and also utilizing some properties obtained by integrating by parts expressions of the type
\begin{equation}
\int_0^1 \!d\xi\ P[\xi^2g+(1-\xi^2)\,H_\star] = P[g] + 2(H_\star-g)\int_0^1 \!d\xi\ \xi^2 P'[\xi^2g+(1-\xi^2)\,H_\star] ~.
\label{byparts}
\end{equation}
Remarkably enough, a neat result comes out after a quite lengthy calculation, once the junction conditions are imposed
\begin{equation}
\widehat{\mathcal{I}}_{bub} = \beta_+M_+ - \beta_-M_- ~,
\end{equation}
which is the exact value needed to correct the on-shell action in such a way that the thermodynamic interpretation is safely preserved. In fact, because of this contribution, the total action takes the --otherwise expected-- form
\begin{equation}
\widehat{\mathcal{I}} = \widehat{\mathcal{I}}_{bh} + \widehat{\mathcal{I}}_{bub}=\beta_+M_+ - S ~.
\label{treintaytres}
\end{equation}
That is, the brane contributes as mass (carrying the mass difference between the two branches) but not as entropy, solely coming from the event horizon of the inner black hole. This is a remarkable result, providing a non-trivial consistency check of the thermalon mechanism. From the Hamiltonian point of view this is naturally understood as follows: Given that the canonical action vanishes, the only possible contributions shall come from boundary terms both at infinity and at the horizon, yielding respectively $\beta_+M_+$ and the entropy, which are nothing but the total charges of the solution.

For the above result we have just imposed the first junction condition, $\widetilde{\Pi}[g_\pm^\star,H_\star]=0$, where $H_\star=\sigma/a^2_\star$ is the static value of $H$. The other junction condition corresponds to the radial derivative of the first one in such a way that
\begin{equation}
\partial_a \widetilde{\Pi}[g_\pm^\star,H_\star]=\partial_{g_+} \widetilde{\Pi}[g_\pm^\star,H_\star] \,\left.g'_+\right|_{a_\star} + \partial_{g_-} \widetilde{\Pi}[g_\pm^\star,H_\star] \,\left.g'_-\right|_{a_\star} + \partial_H \widetilde{\Pi}[g_\pm^\star,H_\star] \,\left.H'\right|_{a_\star} = 0 ~.
\end{equation}
The $H$ derivative of $\widetilde{\Pi}$ played a fundamental r\^ole in the discussion of the thermalon stability, as in (\ref{potstability}). Using the expression (\ref{changePi}) we can show that
\begin{equation}
\partial_{g_\pm} \widetilde{\Pi}[g_\pm^\star,H_\star] = \mp \frac{a_\star}{2\sqrt{f_\pm^\star}}\;\Upsilon'[g_\pm^\star] ~.
\end{equation}
where $f_\pm^\star \equiv f_\pm(a_\star)$, and we used $H_\star - g_\pm^\star = f_\pm^\star/a_\star^2$. For the remaining derivative, after a bit of massage we get
\begin{equation}
\partial_H \widetilde{\Pi}[g_\pm^\star,H_\star] = - \frac12 \int_{\sqrt{f_-^\star}/a_\star}^{\sqrt{f_+^\star}/a_\star}\frac{dx}{x^2}\ \Upsilon'[H_\star-x^2] ~.
\end{equation}
Therefore, we can write the second junction condition as 
\begin{equation}
\frac{M_+}{\sqrt{f_+^\star}} - \frac{M_-}{\sqrt{f_-^\star}} = \frac{4\sigma}{d-1}a_\star^{d-4} \partial_H \widetilde{\Pi}[g_\pm^\star,H_\star] ~.
\label{mjunct}
\end{equation}
Notice that, using (\ref{matchbetas}), the left hand side of this expression can be seen to be proportional to the contribution of the bubble (\ref{Ibrane2}) to the on-shell action (equivalently, the free energy). It vanishes for planar topologies. The junction conditions are also very important to ensure a consistent thermodynamic picture. We have seen in the previous paragraphs that the on-shell action adopts the expected form (\ref{treintaytres}) from the semi-classical approach, where the different quantities, (inverse) temperature and mass, correspond to the expected ones in the solution, {\em i.e.},
\begin{equation}
\beta = \beta_+ ~, \qquad {\rm and} \qquad M = M_+ ~.
\end{equation}
The entropy solely comes from the inner black hole horizon. In addition to this, in order to show the consistency of the thermodynamics, the relations between these quantities and the corresponding thermodynamic potentials must also be the right ones. It is enough to show that, assuming the above values for the mass and the entropy, we recover the correct temperature, {\it i.e.}, $T_+$,
\begin{equation}
T = \frac{dM_+}{dS} = \frac{dM_+}{da_\star}\frac{da_\star}{dr_H}\frac{dr_H}{dS} ~,
\end{equation}
where we have to use the implicit relation between the two parameters, $a_\star$ and $r_H$, provided by the mass of the inner black hole,
\begin{equation}
M_-(a_\star) = r_H^{d-1}\, \Upsilon\left[\frac{\sigma}{r_H^2}\right] ~.
\end{equation}
Equivalently, we can use the property of the inner black hole, $dS = \beta_- dM_-$, so that the inverse temperature is
\begin{equation}
\beta = \frac{dS}{dM_+} = \beta_- \frac{dM_-}{dM_+} ~.
\end{equation}
In order for this to be equal to $\beta_+$, as required, we must verify
\begin{equation}
\beta_+ dM_+=\beta_- dM_- = dS ~,
\label{new1law}
\end{equation}
where the differential is taken with respect to the parameter $a_\star$ or, equivalently, $r_H$. This would imply that the first law of thermodynamics holds true not only for the inner black hole ($\beta_-$ and $M_-$) but for the whole configuration ($\beta_+$ and $M_+$). This can be easily proven using both junction conditions. We derive the first one with respect to $a_\star$, taking into account that, in contrast with the differentiation of the potential with respect to $a$ (not $a_\star$), from which we obtained the second junction condition, the mass parameters $M_\pm$ depend now on the variable. The expression itself depends on $M_\pm$ through $g_\pm^\star$,
\begin{equation}
\Upsilon[g_\pm^\star] = \frac{M_\pm(a_\star)}{a_\star^{d-1}} \qquad \Rightarrow \qquad \left.g'_\pm\right|_{a_\star} = \frac{a_\star\,M'_\pm(a_\star) -(d-1)\,M_\pm(a_\star)}{a^d_\star\,\Upsilon'[g_\pm^\star]} ~.
\end{equation}
The final result is the same as (\ref{mjunct}), with an extra term proportional to the derivatives of the mass parameters,
\begin{equation}
\frac{M_+ - \frac{a_\star}{d-1} M'_+(a_\star)}{\sqrt{f_+}} - \frac{M_- - \frac{a_\star}{d-1} M'_-(a_\star)}{\sqrt{f_-}} = \frac{4\sigma}{d-1} a_\star^{d-4} \partial_H \widetilde{\Pi}[g_\pm^\star,H_\star] ~.
\end{equation}
Therefore, the terms proportional to the mass parameters cancel the right hand side and we get
\begin{equation}
\frac{M'_+(a_\star)}{\sqrt{f_+}} = \frac{M'_-(a_\star)}{\sqrt{f_-}} ~,
\end{equation}
that, after multiplication by $\beta_0$ and use of (\ref{matchbetas}), yields (\ref{new1law}), exactly what we were looking for. We have shown that the first law of thermodynamics holds for the whole bubble configuration whereas the thermodynamic parameters verify the expected relations necessary for the consistency of the thermodynamic approach. Having computed the free energy of this geometry, we can now compare it to the rest of the solutions sharing the same boundary conditions and temperature and thus identify the preferred or classical configuration as the one of lower free energy.

\subsection{Thermalons in LGB gravity}

Before analyzing the free energy of the thermalons and the occurrence of phase transitions, let us describe a bit in more detail the configurations we are dealing with in the simplest case of LGB gravity.  In this particular theory just the EH-brach may display a horizon for planar or spherical black holes, it being  the only possible inner branch. The other (accordingly, outer) branch has long been known to be unstable {\it \`a la} Boulware-Deser (BD), its vacuum contains ghosts in the sense that the kinetic term for gravitons has the wrong sign.  We will be interested in the thermodynamics of the system in the case of {\it wrong} boundary conditions, {\em i.e.}, setting the asymptotics corresponding to this ill-defined branch, with some topology and temperature. This may shed some light on the long-standing question of the fate of the ghosty branch in this theory. For planar and spherical topologies, the thermodynamics corresponding to EH-branch asymptotics is unchanged from what has been discussed. 

In the LGB case there is a single potential that has been presented in (\ref{GBpots1}). In order to define a thermalon configuration, we have to find mass parameters $M_\pm$ such that an equilibrium position, $a_\star$, exists. We have to solve $V(a_\star) = V'(a_\star) = 0$, where we should recall that $g_\pm^\star$ are given implicitly, $g_\pm^\star = g_\pm(a_\star)$, once the branches are chosen in such a way that the inner/outer solution corresponds to the stable/unstable branches. We can write the potential (\ref{GBpots1}) in a more suitable way by reducing the order of the polynomial in $g$,
\begin{equation}
V(a_\star) = \frac{a_\star^{d+1}}{24\lambda (M_+ - M_-)} \left. \left[ (1-4\lambda)\,g^\star + (2+\lambda g^\star) \frac{4 M}{a_\star^{d-1}} \right] \right|^+_- + \frac{\sigma}{2} ~,
\label{V2}
\end{equation}
while its $a$-derivative reads:
\begin{equation}
V'(a_\star) = \frac{a_\star^{d}}{12\lambda (M_+ - M_-)} \left. \left[ (1-4\lambda)(d+1)\,g^\star + [d - 17 + 2 (d-5) \lambda  g^\star] \frac{M}{a_\star^{d-1}} \right] \right|^+_- ~.
\label{dV}
\end{equation}
It is then possible to use these two equations to find the mass parameters $M_\pm$ in terms of $g_\pm^\star$ and $a_\star$, and plug these expressions back into $\Upsilon[g_\pm^\star]$ to get the couple of equations 
\begin{equation}
\Upsilon[g_\pm^\star] = 1 + g_\pm^\star + \lambda \left(g_\pm^\star\right)^2 = \frac{1-4\lambda}{4\lambda} \frac{a_\star^2(d-1)(3+2\lambda\, g_\mp^\star) + 4\lambda(d+1)\sigma}{a_\star^2(d-1) + 2(d-5)\lambda\sigma} ~,
\end{equation}
that can then be solved for $g_\pm^\star = g_\pm(a_\star)$. The general expressions are not particularly enlightening but they simplify quite a lot in the planar case given that the $a_\star$-dependence completely drops out. The bubble radius just remains as the relevant scale.

\begin{figure}[ht]
\centering
\includegraphics[width=0.58\textwidth]{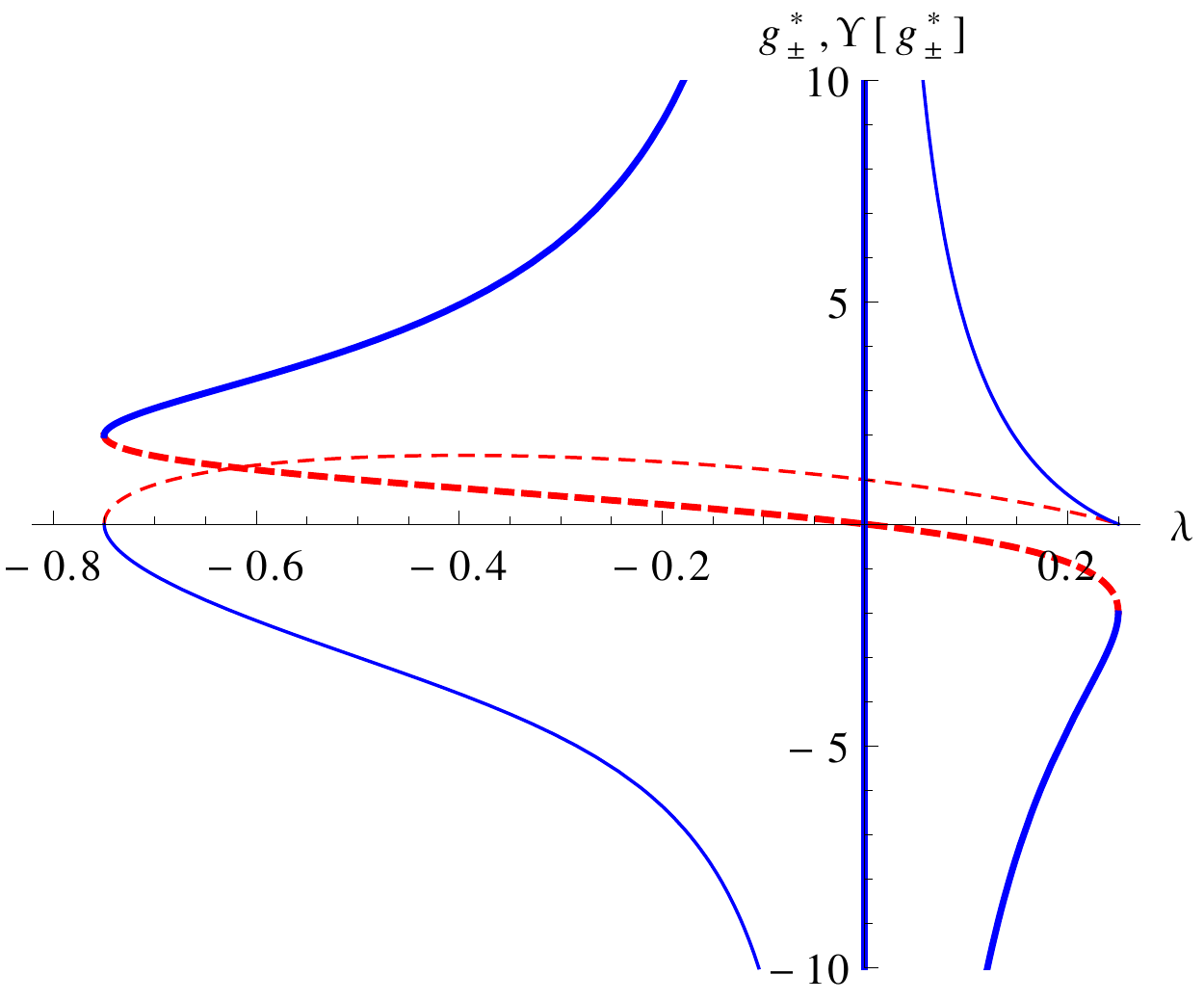}
\caption{$g_{\pm}^\star$ (thick line) and $\Upsilon[g_\pm^\star] \propto M_\pm$ (thin line) are plotted as functions of $\lambda$ in the planar case. Dashed/solid lines corresponding to the inner/outer, $g_-$/$g_+$, branches respectively.}
\label{planargs}
\end{figure}

Let us profit from the simplicity of the planar case to dig a bit further into it. The relevant solution for our purposes is
\begin{equation}
g_{\pm}^\star = \frac{-3+4\lambda\mp\sqrt{3(1-4\lambda)(3+4\lambda)}}{4\lambda} ~,
\label{thermsol}
\end{equation}
since we have to choose the values that correspond to an inner/outer ($-$/$+$) stable/unstable branch. In particular both $g_\pm^\star$ cannot be equal. If we plot these solutions --as well as the mass parameters-- as functions of $\lambda$ (see Figure \ref{planargs}), we can readily see that for negative values of the coupling the outer mass becomes negative, $M_+ < 0$.
Besides, the values of $g_\pm^\star$ are positive for negative $\lambda$ which corresponds to the bubble being formed inside the black hole horizon\footnote{Notice that for $\lambda < 0$ the bubble would separate AdS and dS branches.}, $g_- = 0$. Thus, there are no thermalon configurations for negative values of the LGB coupling. This feature is not specific of planar black holes, no thermalons exist for negative $\lambda$ regardless of the topology. We have already seen this result in the spherical case, in the discussion surrounding Figure \ref{braneHorizon}. For this reason, in the remainder of the paper we will restrict our analysis to positive values of the LGB coupling.

\begin{figure}
\centering
\includegraphics[width=0.58\textwidth]{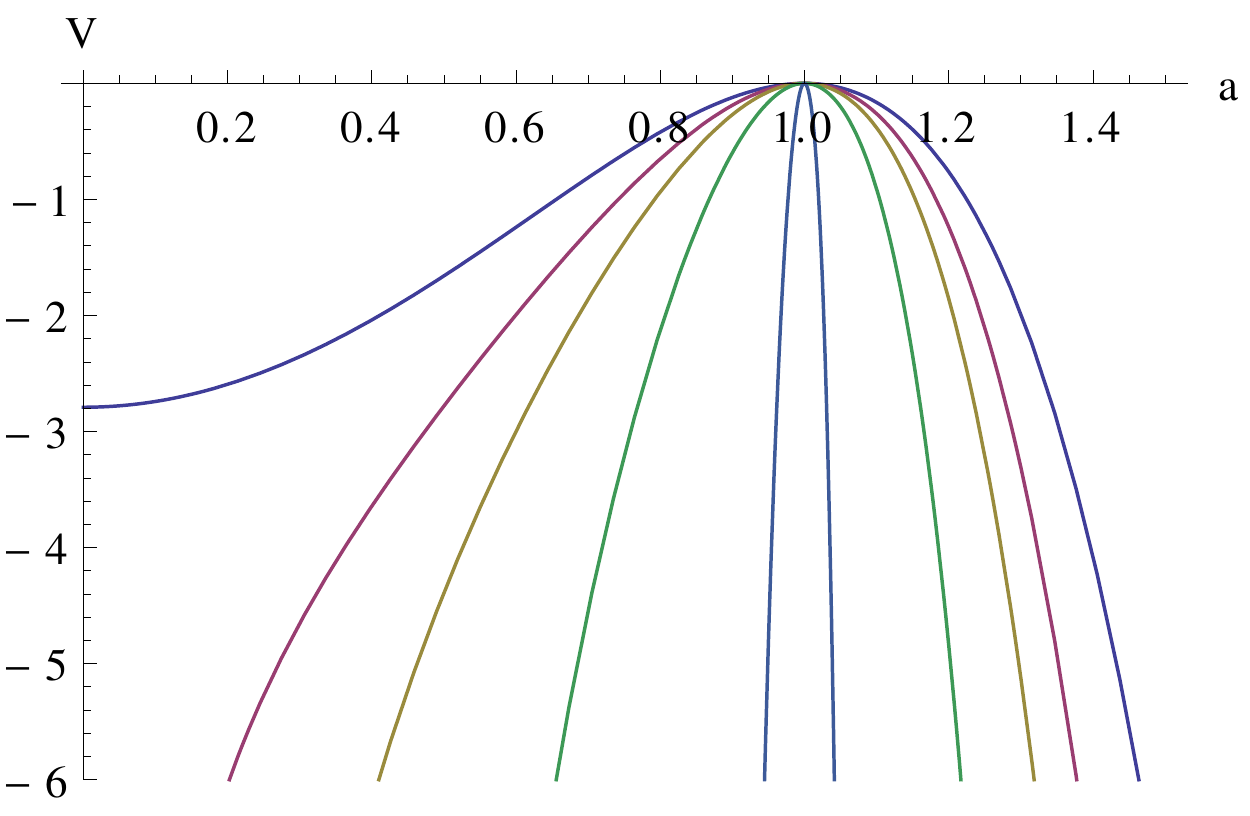}
\caption{Planar bubble potential in LGB gravity for $\lambda=0.1$, $a_\star=1$ and $d=5,6,7,10,50$ (from top to bottom). The non-planar profile is qualitatively the same.}
\label{planarpotential}
\end{figure}

We can plot the potential (\ref{GBpots1}) plugging in the actual values of $M_\pm$ (for a reference value $a_\star = 1$), the qualitative behavior being the same for all values of $a_\star$ and $\lambda$ (positive), even for spherical topology, see Figure \ref{planarpotential}.
Notice the fundamental difference in the behavior displayed by the critical five dimensional LGB gravity, as compared with the higher dimensional theories, in that the potential at zero radius goes to a constant.

In the hyperbolic case the naked singularity of the BD-unstable branch may be covered by an event horizon just for positive $\lambda$ and masses below a certain threshold. This is the only situation in which we may have thermalons with the asymptotics of either of the two branches. Thereby, this is the only setup in which we may have the opportunity of describing transitions between branches occurring in both directions. We will discuss this case in Section \ref{hyperLGB}.

\section{Generalized Hawking-Page transitions in Lovelock gravity}

Having proven the consistency of the thermodynamic picture in the case of the static bubble by deriving (\ref{treintaytres}) and (\ref{new1law}), we are ready to address the question of the global (and local) thermodynamic stability of that configuration. This amounts to analyzing its free energy. The only difference is that we now need to include the thermalons discussed above. This will allow for the study of the formation of these new configurations and their consequences for the dynamics of the system. In the same way as for the usual Hawking-Page phase transition, we can identify the thermodynamically preferred configuration as the background of least on-shell action among all those with a smooth Euclidean continuation and the same period in imaginary time.

\subsection{Planar topology}

The case of planar black holes in general Lovelock gravity can be analyzed straightforwardly. From the point of view of the AdS/CFT correspondence, it corresponds to the familiar situation of CFTs in flat space. The only branch displaying an event horizon is the EH one, all the others being unstable and displaying naked singularities \cite{CamanhoE3}. Because of that, the only possible smooth Euclidean metrics correspond either to the vacua of the theory or to bubble configurations with the EH-branch in the inner region. For any choice of asymptotics (different from EH) we will be comparing, at the same temperature, the corresponding thermal vacuum versus, when available, the thermalon.

The junction conditions simplify a lot in this case, the free energy corresponding just to the black hole with no bubble contribution, as (\ref{mjunct}) made clear. The presence of the junction plays no r\^ole other than setting the temperature difference between the inner/outer regions, according to (\ref{matchbetas}). The free energy of the thermal vacuum is zero, since we have taken it as the reference background, while that of the thermalon yields
\begin{equation}
\mathcal{F}_+ = \frac{\widehat{\mathcal{I}}}{\beta_+} = \frac{\beta_-}{\beta_+} \,\mathcal{F}_- = -\frac{M_+}{d-2} ~,
\end{equation}
which is always negative. This implies that the preferred classical solution is, when available, the thermalon. The most massive one, if there are several with the same asymptotics. For any branch of solutions such that the planar thermalon with inner EH-branch exists, the bubble will always form. Remember that $\beta_- M_- = \beta_+ M_+$ and the inner mass has to be positive in order to have a horizon. In the LGB case, for instance, the thermalon exists for $\lambda > 0$. The equilibrium position being unstable, the bubble will eventually expand engulfing the whole space-time in finite proper time, thereby changing the asymptotics \cite{Camanho2012}.

The existence of these kind of configurations translates into finding $g_\pm^\star$ that verify the static junction equations (\ref{junctionsc}), which, noticing that $H_\star = 0$, can be written as
\begin{equation}
\int_{\sqrt{-g_-^\star}}^{\sqrt{-g_+^\star}} \!dx\ \Upsilon'[-x^2] = 0 ~, \qquad {\rm and} \qquad
\frac{\Upsilon[g_+^\star]}{\sqrt{-g_+^\star}} = \frac{\Upsilon[g_-^\star]}{\sqrt{-g_-^\star}} ~.
\label{staticjunk}
\end{equation}
It is noteworthy that the problem is entirely given in terms of the characteristic polynomial of the relevant Lovelock theory and, as a consequence, as polynomial equations for $g_\pm^\star$. The solutions can be easily plotted and many conclusions can be extracted by analyzing the graphical representation, much along the lines of the methods introduced in \cite{CamanhoE3}.

\begin{figure}[ht]
\centering
\includegraphics[width=0.58\textwidth]{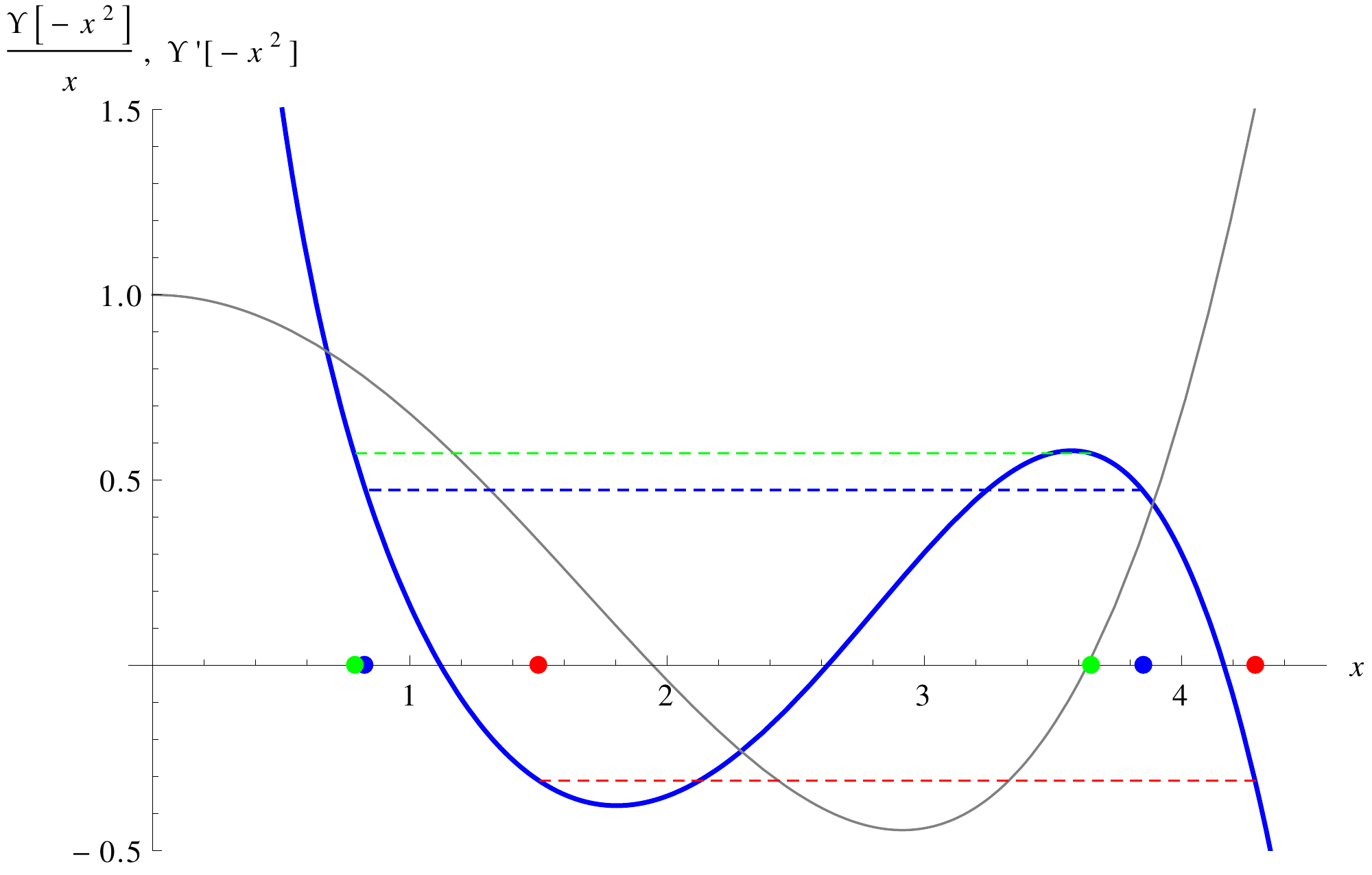}
\caption{$\Upsilon[-x^2]/x$ (thick blue line) and $\Upsilon'[-x^2]$ (thin gray line) for cubic Lovelock theory with parameters $L=1$, $\lambda=0.17$ and $\mu=0.02$. The paired points with the same color --green (left), blue (middle) and red (right) on each branch-- correspond to each of the three solutions of the static junction conditions (\ref{staticjunk}). The area below the gray curve between each couple of points vanishes according to the second equation.}
\label{CubicPlanar}
\end{figure}

Let us illustrate this by using the example provided by a cubic Lovelock gravity, depicted in Figure \ref{CubicPlanar}.
Even though the independent  variable $x=\sqrt{-g}$ is not the usual one for the polynomial, we can easily recognize the three vacua, $\Upsilon=0$, and their associated branches. These correspond to the three ranges with definite sign for $\Upsilon'$, positive for the branches that are stable and negative for the intermediate BD-unstable one. The (stable) EH-branch is the one closest to the vertical axis and then we have two higher-curvature branches. For this particular choice of couplings, we have three possible equilibrium configurations connecting the same two branches. One of the points on each pair corresponds to the EH-branch and its couple belongs to the other stable branch. The blue and green pairs correspond to positive mass solutions whereas the red one has negative mass, both in the outer and inner regions, as indicated by the respective positive and negative values of the polynomial. In the latter case (red dots), the inner region does not display a horizon and thereby the solution has to be discarded. Unlike the LGB case analyzed earlier, there is no thermalon connecting the unstable branch to any of the stable ones for this particular choice of parameters.\footnote{If we insist in fixing such {\it sick} boundary conditions, the system cannot escape via bubble nucleation. Nevertheless, being unstable, it would evolve by some other means, the endpoint of such instability remaining unclear.}

The analysis of cubic Lovelock theory reveals itself particularly interesting since, for the range of couplings displaying three real vacua, it is the minimal scenario where we may analyze the occurrence of bubble transitions between two BD-stable vacua, as in the above example. Notice, however, that if we further demand the absence of causality violation --like the one exhaustively studied in \cite{Brigante,BuchelMyers,Hofman,Boer2010a,CamanhoE1,Boer2010b,CamanhoE2}-- in both branches involved in the transition, we need to go at least to the quartic case (see Appendix). The roots of (\ref{staticjunk}) depend on the values of the coefficients $\lambda$ and $\mu$, though, and we may also have bubbles connecting any of the stable branches with the intermediate unstable one. Bubbles between two higher-curvature branches would never have a smooth Euclidean section, unless the inner region corresponds to the vacuum, something not allowed in general.

We can also analyze the stability of the harmonic potential (\ref{potstability}) at the equilibrium point, whose effective Hooke's constant reads
\begin{equation}
k = \frac{a_\star^2}{2} \left( \int_{\sqrt{g^\star_-}}^{\sqrt{g^\star_+}}\! dx\ \frac{\Upsilon'[-x^2]}{x^2} \right)^{-1}\! \left( \frac{1}{\sqrt{-g^\star_+} \Upsilon'[g^\star_+]} - \frac{1}{\sqrt{-g^\star_-} \Upsilon'[g^\star_-]} \right) ~.
\label{thermstab}
\end{equation}
The positivity of this expression is a standard criterium for stable equilibrium. It is negative for the unique potential of LGB gravity with $\lambda > 0$, as expected for unstable equilibria. This is important for the bubble to rapidly expand and mediate the phase transition. In the cubic case we have two possible potentials, (\ref{newt}), and the equilibrium points can, in principle, be associated to any of them. This seems more involved but one may still use (\ref{thermstab}) in order to study the stability of those static points. For our specific example, we find that the value of $k$ is positive for the red and green pairs, whereas it is negative in the remaining case.
\begin{figure}[ht]
\centering
\includegraphics[width=0.65\textwidth]{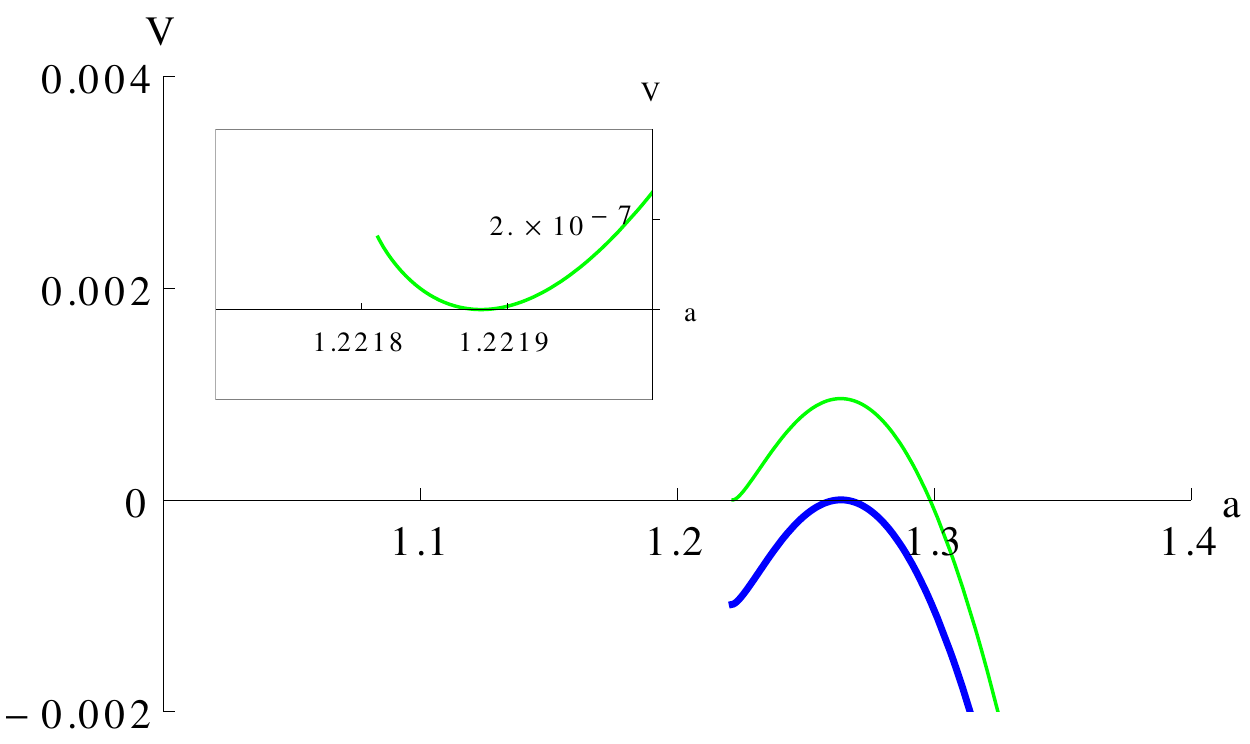}
\caption{Bubble potentials for cubic Lovelock theory with $\lambda=0.17$ and $\mu=0.02$. We just show the relevant branch of the potential connected to the physical equilibrium points, those with positive mass. The upper potential corresponds to the stable configuration (in green in Figure \ref{CubicPlanar}) --notice that the relevant extremum is the minimum where the potential vanishes-- while the other is the unstable one (in blue also in Figure \ref{CubicPlanar}). The origin of this plot corresponds to the horizon of the inner EH geometry.}
\label{cubicPots}
\end{figure}
Thus, we have one possible stable bubble (green) and one unstable (blue) bubble (see Figure \ref{cubicPots}).

We can extract more information from the asymptotic behavior of the potential. For the branch approximated by (\ref{asymptH}), we need first to realize that the outer mass is always bigger than the inner one due to the second condition in (\ref{staticjunk}). We can compute the integral of the polynomial between the two stable vacua which turns out to be negative,
\begin{equation}
\int_{\Lambda_{EH}}^{\Lambda_+}{dx\,\Upsilon[x]} < 0 ~,
\label{polintegral}
\end{equation}
and from (\ref{asymptH}) we can then realize that $H$ is asymptotically negative, which is unphysical.

The bubble cannot reach infinity along this branch of the potential. It can however reach the boundary following the branch that asymptotes to a constant,\footnote{See the discussion at the beginning of Section \ref{fatebubble}.} $H\approx 0.76$, the potential being asymptotically negative along it. This is actually in correspondence with the two positive mass equilibrium points, as seen in Figure \ref{cubicPots}. The unstable bubble may reach infinity by expansion while the other is fixed on its position unless it can {\it tunnel} across the barrier to subsequently expand. The point to the left where the potentials end corresponds to a naked singularity of the outer solution that appears before we have even reached the horizon (situated at the origin of the plot). We expect that the solution becomes unstable before reaching that point, in line with the cosmic censor mechanism discussed in \cite{CamanhoE4}.

Although we have analyzed a very particular example, the same analysis can be performed in general, in cubic or any higher order Lovelock theory. The possible situations one may encounter are extremely varied. As already seen in the cubic case, we may have stable or unstable equilibrium points whose number may change as we vary the values of the couplings, or $a_\star$ for non-planar topology. These may correspond to any of the $K-1$ bubble potentials of the theory, and each of these potentials may have either sign at infinity. This will determine whether the bubble may reach the boundary or not, thus the possibility of a change of branch. Also, depending on the couplings, the branches connected by the static configurations may change; all, some or none being connected to the EH-branch. In addition, for two given branches we may have several static configurations connecting them. All of them have to be compared in order to decide which one is the globally stable phase. Depending on the characteristics of the globally preferred phase for a given asymptotics, the fate of the system may be very different.  The junction conditions considered above determine not only the equilibrium configuration but also, in Lorentzian signature, the effective potential felt by the bubble and, thereby, its subsequent dynamics.

If no thermalon connecting our choice of boundary conditions with the EH-branch exists, the system will remain on the only possible solution, pure thermal AdS$_+$. This is the case for the unstable asymptotics of our cubic example. On the contrary, when the static configuration exists it will form but whether it changes the asymptotics or not will depend on the form of the potential. If the bubble is unstable and no potential barrier appears on its way to the boundary, we will have a phase transition due to a change of branch. This is for instance what happens for LGB gravity with positive $\lambda$.

For a stable bubble the situation is also quite interesting. In a sense, this stable configuration provides a regular black hole to a branch of solutions that naively had none. The bubble being frozen at the equilibrium position, this situation is very similar to the Hawking-Page transition, the system remaining in the black hole phase. In this case, however, when the asymptotic potential allows the bubble to reach the boundary, there might be a non-vanishing probability for the bubble to {\it tunnel} into that region. The boundary conditions may eventually change and we again have a branch transition. This is the case of the stable configuration found in the cubic example (see Figure \ref{cubicPots}).

In case the bubble collapses instead of expanding, it will generally lead to the destruction of the horizon and the consequent formation of a naked singularity. We will not comment more on this here. We will just assume that the system then comes back to the {\it initial} phase, pure AdS$_+$. This is analogous to the formation of bubbles in a fluid. When these expand, the phase transition to the gaseous phase proceeds while, when it collapses, the system remains liquid.

\subsection{Non-planar topology}
\label{hyperLGB}

Even though the equations are much more involved in case of spherical or hyperbolic symmetry, the analysis follows much in the same way. Ultimately, the planar case is a good approximation for the large mass limit of the other topologies. In order to obtain concrete analytic results, we will concentrate in the case of LGB gravity.

The {\it spherical} case was already discussed in \cite{Camanho2012}. We will briefly summarize the results for convenience. The resulting phase diagram is similar to the usual Hawking-Page phase transition. For any (positive) LGB coupling $\lambda$, the free energy of the thermalon configuration, $\mathcal{F}_+$, as a function of the temperature $1/\beta_+$, displays a critical value, $T_c$, above which it becomes negative and, thus, the phase transition occurs. If the free energy is positive, though, the system is metastable. It decays by nucleating bubbles with a probability given, in the semiclassical approximation by $e^{-\beta_+ \mathcal{F}_+}$. Therefore, after enough time, the system will always end up in the stable, EH black hole solution. This is reminiscent of the Hawking-Page transition, except for the fact that the thermal AdS vacuum decays into a black hole {\it belonging to a different branch}. $T_c(\lambda)$ is monotonically decreasing, the phase transition becoming increasingly unlikely when the LGB coupling is turned off.

Asymptotically AdS space-times with {\it hyperbolic} topology are an interesting playground for checking various facts about this novel type of phase transitions. Despite some unclear features of the thermodynamics of these space-times, they allow in particular for the discussion of transitions between two asymptotically AdS branches both having regular black hole solutions. In particular, for the simplest case of LGB theory with $\lambda>0$, both, the EH and the ghosty branch support black holes, the latter depending on the value of the mass. Even though the range of masses for the ghosty branch is bounded from above (and below), it admits black hole solutions for all possible temperatures. Besides, although this branch is unstable, we would like to investigate if this novel transition protects the theory, in the sense that the unstable branch never be the preferred phase, like in LGB theory with planar or spherical symmetry. Despite the fact that the vacuum is preferred for some range of temperatures in the spherical case, the thermalon may be formed with small but finite probability in that case leading to a change of branch.

This case is much richer, not only due to the possibility of transitions in both directions, but also because the spectrum of configurations gets enlarged. The existence of extremal black holes in both branches for specific values of the mass leads to correspondingly extremal thermalons. These extremal configurations can be considered as qualitatively different from their non-extremal counterparts as regularity of the horizon in the Euclidean section does not fix their temperature. As a limit of non-extremal black holes the extremal solutions necessarily have zero temperature and the entropy is given by Wald's formula \cite{Wald}. Quite the opposite, {\it ab initio} extremal configurations may be identified with any temperature and zero entropy \cite{Hawking1995}. The same happens for the bubble configurations even though the limiting temperature is not zero. In addition to the non-extremal bubbles seen so far, there will be extremal solutions where the horizon of the inner black hole is degenerate.

For the direct transition, the one leading to the EH-branch, we will fix the BD-unstable asymptotics and compare the free energy of all possible configurations at the same temperature. These are in principle four families of solutions, extremal and non-extremal black holes and the corresponding bubble counterparts. The same will happen for the reverse transition.

Let us first recall some aspects of the black holes belonging to the unstable branch \cite{CamanhoE3}. These are unstable not only {\it \`a la} Boulware-Deser but also thermodynamically. Their specific heat is negative and they furthermore have negative entropy. As a consequence, their free energy will always be bigger than the one corresponding to the extremal black hole with the same mass. In the critical five dimensional case, the extremal black hole is the most massive, $M = \lambda$, its degenerate horizon coinciding with the singularity at $r=0$. This does not represent a problem since that point is at the end of an infinite throat and should be removed from the geometry. The near horizon geometry can be written as 
\begin{equation}
ds^2 \sim 2 \lambda d\rho^2 + e^{2\rho} \left( - \frac{dt^2}{2\lambda} + d\Sigma_{-1,3}^2 \right) ~,
\end{equation}
where we have explicitly removed the point $r=0$ by considering a change of variables as $\rho = \log r$. In higher dimensions, the extremal black hole is the one displaying a degenerate horizon at finite radius, and its geometry is completely regular. We will concentrate in the more interesting five dimensional case in what follows.

The extremal black hole solution is a smooth geometry interpolating between $AdS_5$ at spatial infinity and the previous geometry as we approach $r=0$. It has zero specific heat and entropy, and can be considered with any periodicity in Euclidean time. Thus, it is a well motivated candidate to be taken as a ground-state of the theory for the sector we are considering. Also, as the black holes are unstable, all of them have higher free energy than the extremal one and thus the chosen ground-state is always preferred in a semiclassical basis. The only relevant configuration for the next step of the analysis will then be the extremal black hole.

\begin{figure}[ht]
\centering
\includegraphics[width=0.39\textwidth]{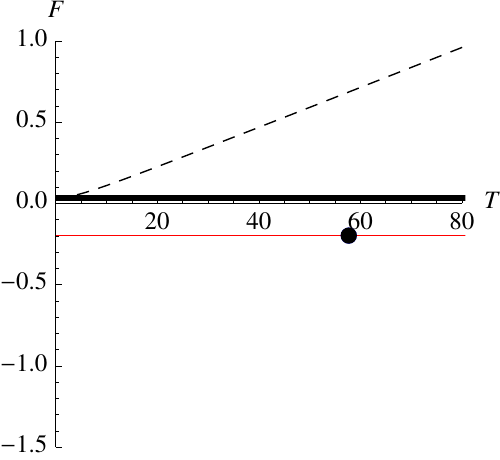}\quad\includegraphics[width=0.39\textwidth]{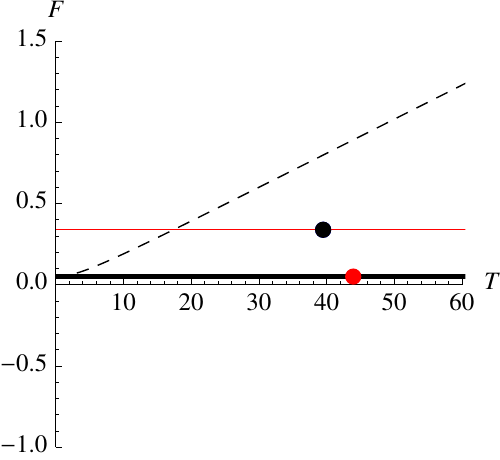}\\[1em]
\includegraphics[width=0.39\textwidth]{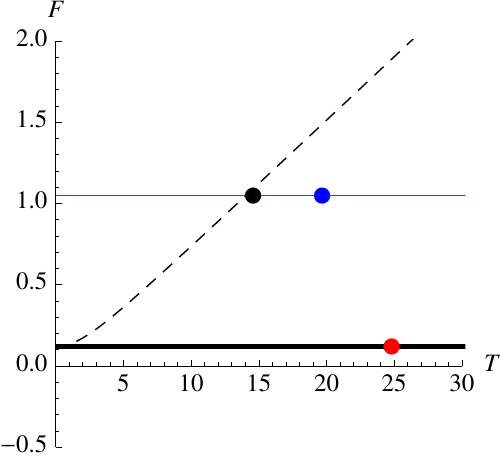}\quad\includegraphics[width=0.39\textwidth]{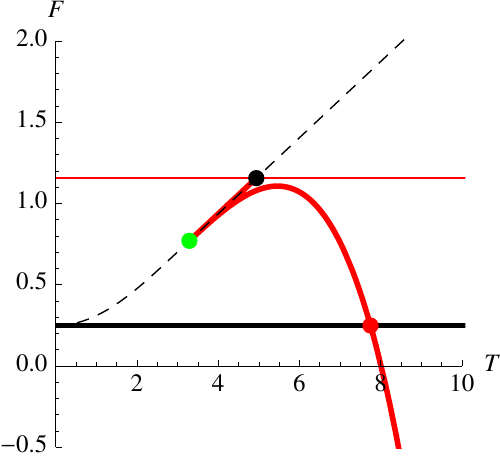}
\caption{Free energy versus temperature for $\lambda=0.035, 0.05, 0.12, 0.249$. The dashed black curve corresponds to the unstable branch of black holes while the thick black one represents its extremal free energy. The thick red curve (in gray in the printed version) corresponds to the bubble with the same asymptotics and its extremal limit is indicated by the black dot. The thin red curve corresponds to the free energy of such extremal state when vanishing entropy is assumed. The dots indicate the locus of different kinds of possible phase transitions even though just some of them may take place in each case. The black dot when relevant represents a phase transition from an extremal to a non-extremal bubble, and the red one a transition from the extremal black hole to the non-extremal bubble. The green dot (in lighter gray) indicates the minimal temperature for the existence of non-extremal bubbles when these are divided in two branches, one thermodynamically stable and one unstable.}
\label{directH}
\end{figure}

We will have to compare the free energy of our thermalon configurations to that of this ground-state. The free energy of the non-extremal configuration has the usual expression, $\mathcal{F} = M - T S$, while the one corresponding to the extremal one, once again corresponds to a constant thus implying a vanishing entropy. However, the extremal free energy does not coincide with the mass in this configuration. In the extremal case, the bubble is situated exactly at the black hole horizon of the inner region, $f_-(a_\star)=0$, in such a way that the rescaling of the temperature just cancels the zero of $f'$,
\begin{equation}
\frac{\sqrt{f_+(r_H)}}{\sqrt{f_-(r_H)}}\, \frac{f'_-(r_H)}{4\pi} \to \tilde{T}_+^{\,e} ~.
\end{equation}
This is the temperature corresponding to the black dot where the non-extremal bubble curve ends, in Figure \ref{directH}.
The resulting on-shell action is then, $\widehat{\mathcal{I}}^{\,e} = \beta^{\,e}_+ (M_+^{\,e}-\tilde{T}_+^{\,e} S^{\,e})$. The free energy corresponds to an {\it effective} mass that picks some contribution proportional to the Wald entropy,
\begin{equation}
\mathcal{F}^{\,e}_{bub} = M^{\,e}_+ - \tilde{T}^{\,e}_+ S^{\,e}
\end{equation}
Here the word extremal does not necessarily mean zero temperature but it is rather a question about the topology of the near horizon region. For the discussion of the phase diagram we will assume that the extremal configurations are present at any temperature with zero entropy. At the end of this Section we will comment on the alternative approach, not considering them but as limiting cases  of their non-extremal counterparts.

The curves for the free energy as a function of the temperature for the different configurations are depicted in Figure \ref{directH} for several values of the coupling $\lambda$. We have different behaviors for different ranges of the LGB coupling. For small values, $\lambda < \lambda_{crit} \approx 0.0404$, the bubble is always the stable phase with a transition between the extremal configuration at low temperatures and the non-extremal one, which is preferred as we increase the temperature. For values of $\lambda$ above $\lambda_{crit}$ the low temperature phase is, instead, the extremal black hole, our hypothetical vacuum, the non-extremal bubble being again the stable phase at high temperature. Another interesting feature is that for values of $\lambda$ bigger than $1/12$, bubble configurations with negative entropy appear, as noticeable from the positive slope of the non-extremal curve in the third and fourth graphics. This negative entropy states, though, are never the preferred phase of the system and can safely be discarded as unphysical.

A qualitatively different scenario would have emerged if we considered the alternative approach to the extremal states mentioned earlier in which these are just limiting cases of the non-extremal configurations already included as endpoints of the corresponding curves (black dot for the extremal bubble). We would have to consider the same graphics shown in Figure \ref{directH}, but discarding the thick black and thin red curves corresponding to the extremal states. The stable phase at low temperature would then always be the black hole, with bubble formation at high temperature. This situation is puzzling as the low temperature phase would have negative specific heat and entropy. The negative entropy bubble states would also be the preferred phase for some range of the temperature in that case. We would be unable to just discard them as unphysical as we would not have any available metric at low temperatures. The inclusion of the extremal states seems to solve the problem of negative entropy configurations, reducing it to a matter of thermodynamic instability.

\begin{figure}[ht]
\centering
\includegraphics[width=0.58\textwidth]{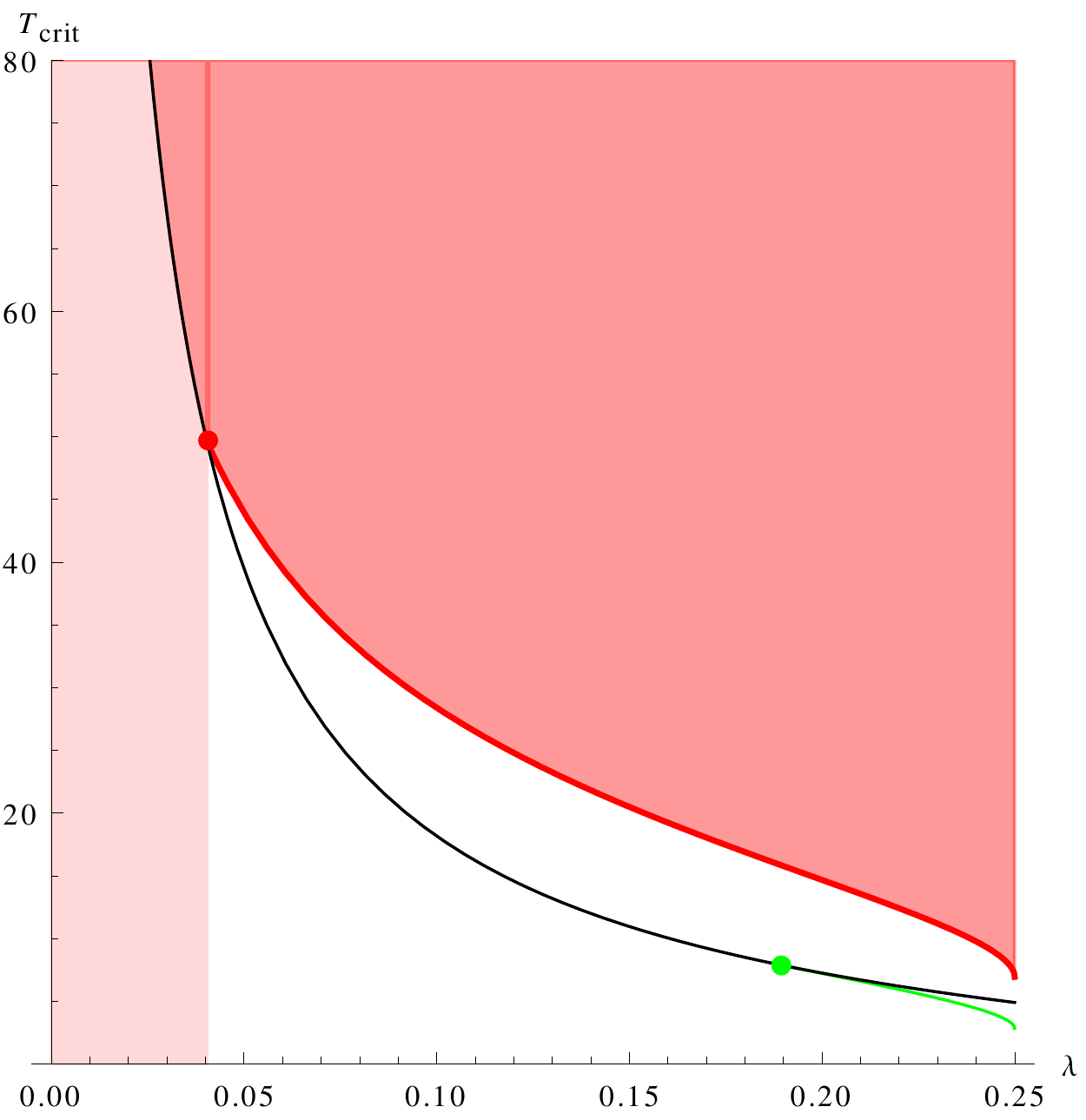}
\caption{In red (gray in the printed version) the bubble states, light for the extremal, and in gray the extremal (BD-unstable) black hole. The red dot indicate the value of $\lambda$ for which the transition between the extremal unstable black hole and the non-extremal bubble (thick red line) begins to exist. The black line corresponds to the bubble extremal limit and the green one (lighter gray) to the minimal temperature of these configurations when this does not correspond to the extremal one. The colors of the different lines are the same used in Figure \ref{directH} to indicate the points in the plot those lines represent.}
\label{directHT}
\end{figure}

In the region of the $\lambda-T$ phase diagram of Figure \ref{directHT} colored in white, the bubble configurations have higher free energy than the {\it vacuum}, this being the thermodynamically preferred phase.
Still, the probability of bubble formation is non-zero, being proportional to the exponential of the difference of the actions of both solutions. Thus, after enough time a bubble will form leading again the system to the EH-branch.

\begin{figure}[ht]
\centering
\includegraphics[width=0.39\textwidth]{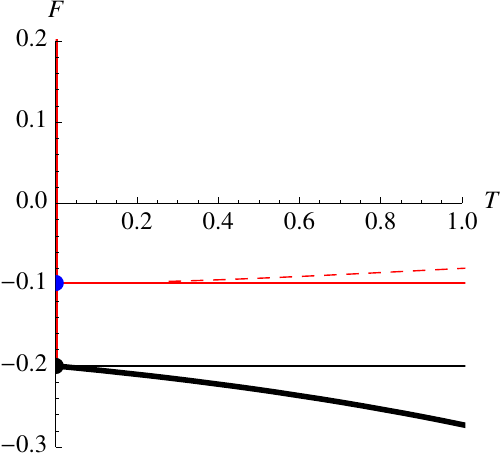}\qquad\includegraphics[width=0.39\textwidth]{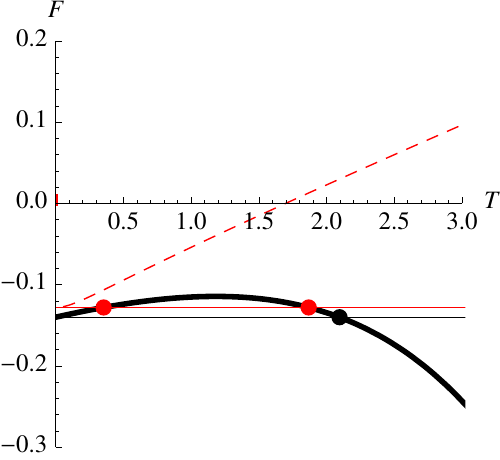}\\
\includegraphics[width=0.39\textwidth]{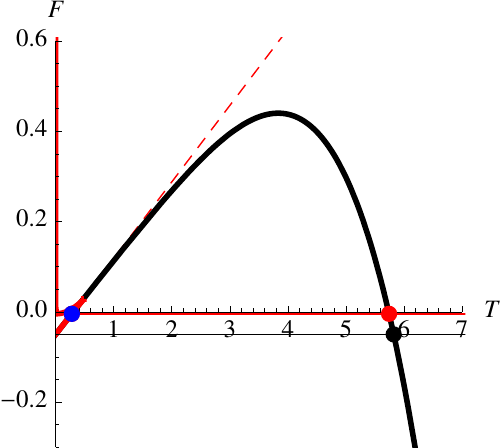}\qquad\includegraphics[width=0.39\textwidth]{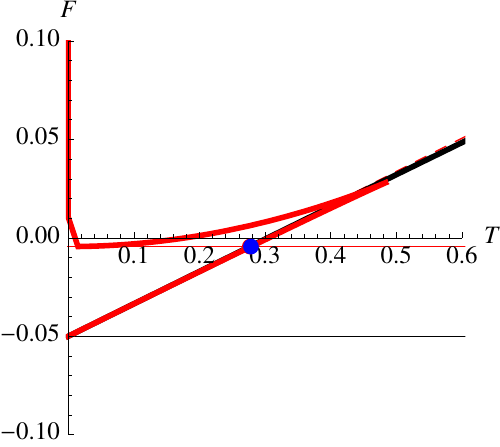}
\caption{Free energy versus temperature for $\lambda=0.05,0.11,0.2 \,\text{(and zoom)}$. In red (gray in the b$\&$w version), the bubble states and in black the black holes in the stable branch. The thin black line corresponds to the extremal geometry once assumed the vanishing of the entropy. The red (gray) dots indicate the locus of extremal bubble to black hole transitions and viceversa. The dashed (red) line corresponds to the extension for $a < \frac{1}{\sqrt{2}}$ of $\mathcal{F}_+$. These are not viable bubble solutions in the same way as its extremal extension represented by the thin red (gray) line. Acceptable bubble geometries appear for $\lambda \geq \lambda^\star$ (see the two lower figures) represented by the thick red line and they also have a well defined extremal extension.  The blue dot indicates non-extremal to extremal bubble transitions.}
\label{inverseH}
\end{figure}

For the {\it inverse} transition we proceed in the same way but setting the opposite asymptotics. The results are depicted in Figure \ref{inverseH}.
Strikingly, due to several physical constraints, for low values of the LGB coupling bubbles --either extremal or non-extremal-- do not exist, the critical value being $\lambda^\star = \frac{1}{24}(7-\sqrt{13}) \approx 0.14$ where the two extremal states degenerate. They are displayed in the Figure (dashed red and thin red respectively for non-extremal and extremal bubbles) but they would be formed inside the outer horizon. Viable bubble geometries appear above the critical $\lambda$ value represented by the thick red line and they also have a well defined extremal extension (thin red line). 

For low enough values of the LGB coupling the black hole states are the only available configurations. Below $\lambda=1/12$ the non-extremal solutions are stable and have positive entropy thereby having lower free energy than the extremal {\it vacuum}. The non-extremal black hole is the stable phase of the system for all temperatures. For higher values of $\lambda$, negative entropy non-extremal states appear at low temperatures but the extremal configuration has lower free energy in that case. We can again safely remove the unphysical states. The low temperature phase corresponds to the extremal black holes with a transition to the non-extremal ones at high temperature.

The stable low  and high temperature phases are still the same above the critical value $\lambda^\star$, where the bubble configurations appear. These can be divided in two branches, one stable that merges with the black hole curve at zero temperature and one unstable close to the extremal state that is irrelevant, having higher free energy than the former. Any of these bubbles have higher free energy than the extremal black holes. Still, and as for the transition in the other sense, the probability of a bubble being formed is non-zero and thus after enough time a bubble of the BD-unstable phase would form. This would in principle drive the system towards the pathological branch but it is also unstable to the formation of bubbles of the EH-branch. The final fate of the system seems to be a chaotic situation with bubbles of both phases popping up everywhere. This seemingly problematic situation is avoided for $\lambda < \lambda^\star$. It may be convenient to recall at this point that the LGB coupling is subjected to stronger constraints if the theory is to prevent causality violation \cite{Brigante}.

\begin{figure}[ht]
\centering
\includegraphics[width=0.55\textwidth]{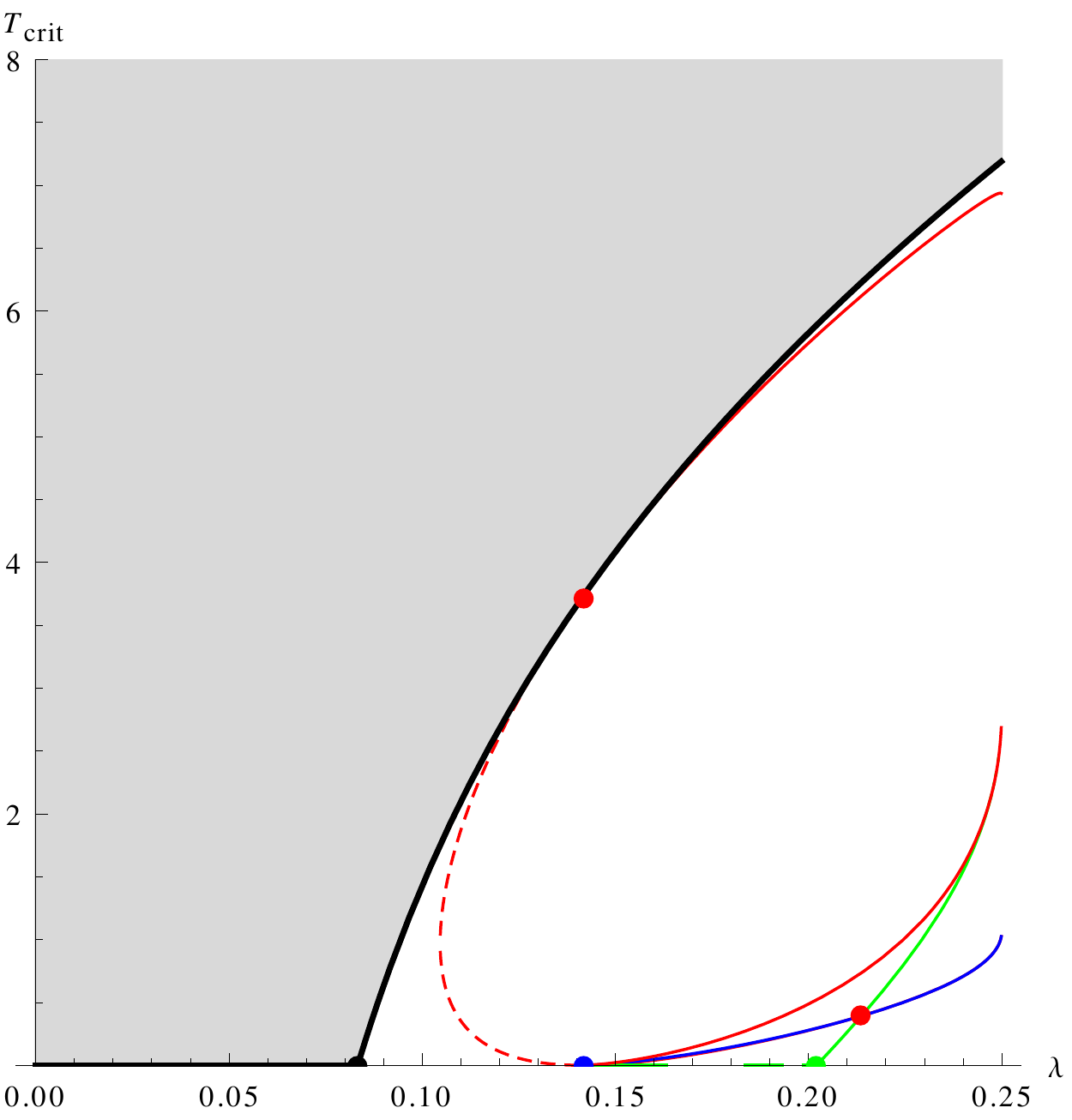}
\caption{The phase diagram is divided in two regions in this case both corresponding to black holes, either extremal (white) or non-extremal (gray). The thick black line corresponds to a extrema to non-extremal phase transition, the rest of the lines representing would be transitions involving bubbles. The dots where the dashed line ends indicate the value of lambda for which such bubble configurations actually exist. The colors of the different lines are the same used in Figure \ref{inverseH} to indicate the points in the plot those lines represent.}
\label{inverseHT}
\end{figure}

If we choose to discard the free energy curves corresponding to the extremal states, thin red and black lines, we would encounter problematic behavior for even lower values of $\lambda$. For $\lambda > 1/12$, negative entropy black holes would appear and they would be the preferred phase for low temperatures. Above $\lambda^\star$, the bubbles appear --also with negative entropy-- with lower values of the free energy for some range of temperatures, between the lower thin red and green curves of Figure \ref{inverseHT}. 

In summary, for values of the LGB coupling below $\lambda^\star$ the BD-unstable branch of hyperbolic black holes is always driven to the EH one, this being stable against the formation of bubbles. For this scheme the EH-branch is protected and the only transition that takes place is a change from the extremal black hole at low temperatures to the non-extremal at higher ones. This transition only occurs when negative entropy states are in the spectrum, namely close to the extremal configuration. For higher values of $\lambda$ the situation is chaotic with bubbles of the opposite branch popping up for any choice of the boundary conditions, the interpretation of this being unclear. This might be seen as another hint pointing toward inconsistencies of the gravitational theory for such large values of the higher-curvature couplings. The results of this paper suggest that, for small enough values of $\lambda$, the LGB theory rejects being forced to the BD-unstable branch by means of a phase transition towards the EH-branch. This will also be presumably valid in the generic Lovelock theory, for values of the couplings for which the theory is healthy. By this we mean that features like causality violation or generic linear instabilities are absent \cite{Camanho2010d}. This may be an explanation for the distinctive r\^ole played by the EH-branch in higher-curvature theories of gravity.

\section{Discussion}

In this paper we have broaden the scope of our analysis of \cite{Camanho2012} about phase transition in Lovelock theories. That is, we have studied the possibility of distributional solutions for which the spin connection is discontinuous at some given junction (hyper)surface. In the absence of matter in the {\it bubble}, it glues two solutions that correspond to different branches of the same theory, {\em i.e.}, to different asymptotics. From the Hamiltonian point of view, the existence of such configurations is allowed by the multi-valuedness of the canonical momenta, a peculiar dynamical property of these models. 

We have proven that it is possible to generalize the thermodynamic notions usually applied to black holes to comprise these new solutions. Under certain regularity assumptions, static bubble configurations can be assigned temperature, mass, entropy and free energy, all verifying the expected thermodynamic relations. Having proven the consistency of the thermodynamic picture, we have then  analyzed local and global stability of our system in this generalized context and the occurrence of phase transitions. We have dedicated special attention to LGB gravity even though the same kind of transitions occur also in the general case, as it can be explicitly shown for planar topology. Besides, in a higher-order case -say cubic gravity action- new phenomena like the existence of more than one type of bubble appear. This kind of effect was not present in the simpler example worked out in \cite{Camanho2012}. The bubble nucleation described here is a novel mechanism for phase transitions that is a distinctive feature of higher-curvature theories of gravity. Specifically, phase transitions among the different branches of the theory are driven by this mechanism. Mimicking the thermalon configuration \cite{Gomberoff2004}, a bubble separating two regions of different cosmological constants pops out, generically hosting a black hole.

In the context of LGB gravity, this configuration is thermodynamically preferred above some critical temperature. The corresponding phase transition can be interpreted as a generalized Hawking-Page phase transition for the high-curvature branches, even if they do not admit black holes themselves, driving the system towards the EH-branch. This happens even for the hyperbolic case in which the reverse transition is also possible. This is a novel feature whose interpretation is unclear to us, in the sense that it does not seem a natural behavior for the system. It may be pinpointing to the necessity of narrowing the range of admissible higher-curvature gravitational couplings. For the EH-asymptotics the usual black hole is always the preferred phase. Below some critical value of the LGB coupling $\lambda$, thermalon solutions do not even exist in this sector of the theory. 

The junction conditions do not just determine the existence of the static configurations but also their dynamics. In the LGB case, the bubble configuration, being unstable, dynamically changes the asymptotic cosmological constant, transitioning towards the stable {\it horizonful} branch of solutions, the only one usually considered as relevant. This is then a natural mechanism for the system to select the general relativistic vacuum among all possible ones. We are aware of the fact that the vacuum $\Lambda_+$ in the LGB theory exhibits ghosts. However, the phenomenon takes place, for instance, in the cubic Lovelock theory as well, where there are further healthy vacua than the one connected to the EH action \cite{Camanho2010d}.

The selection of the EH-branch among all the stable ones is not as universal as one may naively think, though, as not all branches of solutions are connected to it by a thermalon. Thus, choosing such boundary conditions the asymptotics would not change due to the mechanism presented here. Besides, even when a suitable static configuration exists, the dynamics of the bubble might not allow the bubble to escape to infinity, {\em e.g.}, those cases in which the equilibrium position of the bubble is stable, so that it cannot change the asymptotics either. In those cases, bubble configurations might represent new phases for higher-curvature vacua, in some situations very similar to a regular black hole. Remember that in many cases these branches do not admit black hole solutions. 

Usually, AdS space-times are considered to have perfectly reflective boundaries so that any excitation bounces back in finite proper time. We have taken a different approach in this paper. As the theory has several possible asymptotics, all {\it \`a priori} equally valid, we might choose any of them but we have to allow for asymptotics changing solutions such as those described in this paper. Through the evolution of the bubble, these also change the temperature and the mass of the solution, as it happens in the case of {\it quenches} (see \cite{Buchel2013a} for an example in the context of holography). We might think of the thermalon mechanism as a kind of {\it thermodynamic quench} induced by the system, not driven by external means.

Let us conclude by saying that, following the same approach that led us to consider the thermalon configurations discussed so far, one can include in the analysis more sophisticated solutions. For instance, in the LGB case with positive $\lambda$ (see Figure \ref{rootchange}), for values of $M_+$ slightly above the thermalon value giving the equilibrium point, the Euclidean trajectory would be oscillatory, between a maximal and a minimal value of the radius, $a_\pm$. The same happens for any unstable bubble. Instead of picking the static value, we may tune the value of $M_+$ so that the period of oscillation
\begin{equation}
\beta=2\int_{a_-}^{a_+}\frac{\dot{T}_-}{\sqrt{2 V(a)}}da ~,
\end{equation}
{\it fits} the periodicity $\beta_-$ given by the regularity at the horizon, an integer number of times.
\begin{figure}[ht]
\centering
\includegraphics[width=0.58\textwidth]{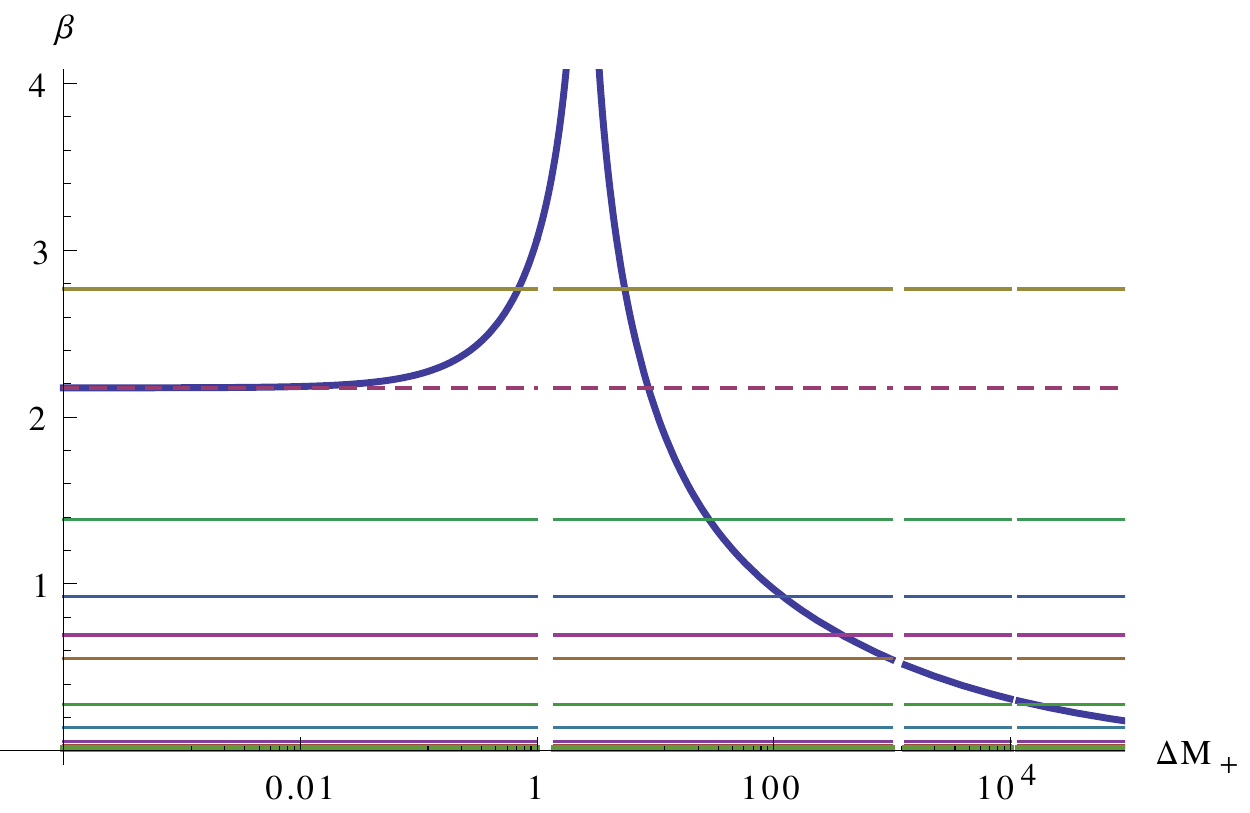}
\caption{Oscillation period for a bubble in 5d LGB gravity for $\lambda=0.1$, $M_- = M_-^{th}$ and $M_+ = M_+^{th} + \Delta M_+$, $M_\pm^{th}$ being the mass parameters corresponding to $a_\star=1$. The horizontal lines, from top to bottom, correspond to $\beta_-/n$ for $n=1,2,3,4,5,10,20,50$. In this specific example we have an infinite tower of smooth manifolds with increasing $n$ and $M_+$.}
\label{ISpectrum}
\end{figure}
We can plot this periodicity as a function of $M_+$ (see Figure \ref{ISpectrum}) and identify the values verifying
\begin{equation}
\beta_- = n \beta ~,
\end{equation}
for some integer $n$, as the values of the mass parameter that yield smooth Euclidean manifolds. The picture is similar to the thermalon of Figure \ref{thermalon} with a wiggly line as a junction. The above condition is such that the trajectory is closed. These new solutions have to be considered as well in the canonical ensemble and may lead to more general phase transitions than those previously discussed.

A possible interpretation of the mechanism described in this paper is that of it revealing a new type of instability of Lovelock theory, which seems to occur even in the case of vacua that in principle are free of other pathologies like BD-instability. On the other hand, the phenomenon described here may also be thought of as a dynamical process that can be interpreted in terms of AdS/CFT correspondence. That is, Lovelock gravities, as well as any other higher-curvature theory, have several branches of asymptotically (A)dS solutions that might admit an interpretation as different phases of the dual field theory. Phase transitions among these are driven by the mechanism described in the present paper. One might also feel tempted to argue that, from the holographic point of view, the phenomenon observed here looks like a confinement/deconfinement phase transition in a dual CFT, involving an effective change in the microscopic properties ({\em e.g.}, the central charges), both phases being strongly coupled. Whether a phenomenon like this takes place in a 4d or higher dimensional CFTs, particularly within the framework of the fluid/gravity correspondence --where both phases might be characterized by different transport coefficients--, or it is overtaken by higher-curvature corrections, is an open question at this point. To scrutinize whether this is the case or, else, these pathologies should be interpreted as tighter physical constraints on otherwise admissible higher-curvature gravities, is still an open issue.

\section*{Acknowledgements}

These results were presented in a number of talks over the last months. We would like to thank those who enriched the content of this manuscript through fruitful questions and discussions.
The work of X.O.C. and J.D.E. was supported in part by MICINN and FEDER (grant FPA2011-22594), by Xunta de Galicia (Conseller\'{\i}a de Educaci\'on and grant PGIDIT10PXIB206075PR), and by the Spanish Consolider-Ingenio 2010 Programme CPAN (CSD2007-00042).
The work of G.G. was supported by NSF-CONICET, PIP, Math-SUD, BMWF-MINCyT, PICT, and UBACyT grants from CONICET, ANPCyT and UBA.
The work of A.G. was partially supported by Fondecyt (Chile) Grant $\#$1090753.
X.O.C. and J.D.E. would like to thank the FCEN-UBA and UNAB for hospitality during part of this project.
X.O.C. is thankful to the Front of Galician-speaking Scientists for encouragement.
The Centro de Estudios Cient\'{\i}ficos (CECs) is funded by the Chilean Government through the Centers of Excellence Base Financing Program of Conicyt.

\appendix
\section{Causality for higher curvature vacua}

Following the line of research opened in \cite{Brigante} in the case of quadratic gravity, the issue of causality has been addressed in the context of Lovelock gravity in \cite{Camanho2010d}. Analyzing the propagation of gravitons on the background of a planar black hole, it has been found that the Lovelock couplings must verify a set of inequalities, that in turn ensure the positivity of energy correlators in the holographic dual \cite{HofmanMalda}. The one-point correlator, that can be given in terms of two- and three-point functions of the energy momentum tensor, is fixed up to two constants, $t_2$ and $t_4$, that have been computed holographically yielding
\begin{equation}
t_2=-\frac{2(d-1)(d-2)}{(d-3)(d-4)}\frac{\Lambda \Upsilon''[\Lambda]}{\Upsilon'[\Lambda]} ~, \qquad t_4 = 0 ~,
\end{equation}
where $\Lambda$ is the effective cosmological constant of the vacuum we are interested in. In accordance with this, we get different constraints on the couplings for the different branches.

These causality constraints are usually discussed for the EH-branch of the theory. It is the only one with black hole solutions in the planar case, thus the only one for which the black hole computation we were referring to is valid. Nonetheless, one may still wonder which constraints imposes positivity of the energy for the CFTs defined by these vacua. Notice that the values of $t_2$ changes if we focus in a different vacuum whereas $t_4$ remains zero. A different computation using a shock wave exact solution provides the corresponding violation of causality for all branches \cite{Hofman}.  

Causality constraints can be rephrased as constraints in the characteristic polynomial about the specific vacuum under consideration. For the EH-branch, it is expected that for any Lovelock theory of arbitrary degree and spacetime dimensionality there is always a region containing the Einstein-Hilbert point where causality (and stability) is respected. These constraints can be rewritten for BD-stable AdS vacua as
\begin{equation}
-\frac{(d-3)(d-4)}{2(d-1)} \leq C[\Lambda]\equiv\frac{\Lambda\Upsilon''[\Lambda]}{\Upsilon'[\Lambda]} \leq \frac{(d-3)}{2(d-1)} ~.
\end{equation}
The upper and lower bounds correspond to the zero helicity or scalar channel and the helicity two or tensor channel respectively. The helicity one condition is always less constraining and can be safely ignored. Let us remark that these bounds depend on the dimensionality whereas every other expression involved in the problem does not (once fixed the degree of the polynomial). The upper bound is itself bounded from above as
\begin{equation}
\frac{(d-3)}{2(d-1)}<\frac12 ~,
\end{equation}
while the lower bound is unbounded. Furthermore, we may write the polynomial in terms of its roots
\begin{equation}
\Upsilon[\Lambda] = c_K \prod_{k=0}^K (\Lambda-\Lambda_k) ~.
\label{pol2}
\end{equation}
From the reality of the Lovelock coefficients (or equivalently of the Lagrangian) we know that the roots are either real or come as conjugate dual pairs. The only further constraint we will consider is that it has a well defined (AdS-)EH limit, {\it i.e.},
\begin{equation}
\Upsilon'[g] > 0 ~,\qquad \forall g\in [\Lambda_{EH},0) ~.
\end{equation}
This defines effectively the EH vacuum. From the characteristic polynomial (\ref{pol2}) we can easily express $C[\Lambda_i]$ for each vacuum as
\begin{equation}
C[\Lambda_i] = 2 \sum_{j\neq i} \frac{1}{1-\frac{\Lambda_j}{\Lambda_i}} .
\label{C}
\end{equation}
Notice that also the BD-stability of such vacuum can be very easily assessed in this way, as the first derivative of the polynomial there yields
\begin{equation}
\Upsilon'[\Lambda_i] = c_K \prod_{j\neq i} (\Lambda_i-\Lambda_j) ~.
\end{equation}
Starting from the stable EH-vacuum, the AdS roots follow a series of stable/unstable vacua where complex conjugate pairs of roots do not matter. The same happens for the dS vacua. For $\Lambda_i<0$, we can classify the contributions to (\ref{C}) according to their sign
\begin{equation}
1 - \frac{\Lambda_j}{\Lambda_i} \geq 0 \qquad \Longleftrightarrow \qquad \Lambda_i \leq \Lambda_j ~,
\end{equation}
and the reverse for the opposite sign. This is also valid for the complex conjugate pairs where $\Lambda_j$ has to be replaced by its real part. 

Using this it is very easy to show that in cubic Lovelock gravity, the minimal example with two BD-stable AdS-vacua, only the EH-vacua may be causal. In the case in which both such vacua exist, the characteristic polynomial displays three real roots 
\begin{equation}
\Lambda_{HC} < \Lambda_{u} < \Lambda_{EH} < 0 ~,
\end{equation}
where the subscripts stand for higher-curvature, unstable and Einstein-Hilbert respectively. If we want to scrutinize if causality holds for the higher-curvature vacuum, $\Lambda_{HC}$, we just need to analyze two terms, both positive and bounded from below
\begin{equation}
C[\Lambda_{HC}] = \frac{2}{1-\frac{\Lambda_u}{\Lambda_{HC}}} + \frac{2}{1-\frac{\Lambda_{EH}}{\Lambda_{HC}}} > 4 ~,
\end{equation}
in such a way that $C[\Lambda_{HC}]$ automatically violates the helicity zero constraint. The same happens, in general, for the smaller root of the polynomial, when real. Assuming that there is at least one real AdS root (for instance the EH one) and that the rest have bigger real part, we can easily show that
\begin{equation}
C[\Lambda_{smaller}] > 2 ~.
\end{equation}
In case there is any smaller root --with more negative real part-- this would yield a negative contribution to $C[\Lambda_i]$. In absolute value, this term may be as big as one wants, the corresponding (unstable) vacuum just has to approach sufficiently the one we are analyzing. Thus, one negative term alone may compensate any positive contribution from the rest of the roots and bring the value of $C[\Lambda_i]$ to the interval allowed by causality.

The minimal example for which we may have a BD-stable and causal AdS vacuum in addition to the EH one is quartic Lovelock theory, we just need $c_4>0$ in the region of parameter space where the four roots are real and negative.


\bibliographystyle{plain}

\end{document}